# HYPERSURFACE CURVATURES OF GEOLOGICAL FEATURES

*Igor Ravve, Anne-Laure Tertois, Bruno de Ribet, and Zvi Koren*

*Aspen Technology, Houston, Texas, USA*

Igor.Ravve@aspentech.com, Anne-Laure.Tertois@aspentech.com,

Bruno.deRibet@aspentech.com, Zvi.Koren@aspentech.com

## ABSTRACT

Reflector-normal angles and reflector-curvature parameters are the principal geometric attributes used in seismic interpretation for characterizing the orientations and shapes, respectively, of geological reflecting surfaces. Commonly, the input dataset for their computation consists of fine 3D grids of scalar fields representing either the seismic-driven reflectivities (e.g., amplitudes of 3D seismic migrated volumes) or model-driven reflectivities, computed, for example, from the derived elastic impedance parameters. Conventionally, the computation of curvature parameters at each grid point is based on analyzing the local change in the inline/crossline dips, considering the potential existence of a local quadratic reflecting surface in the vicinity of that point. This assumption breaks down for subsurface points in the vicinity of either complex reflecting surfaces (e.g., brittle/rough/tilted synclines/anticlines, ridges/troughs, and saddles) and/or sharp, discontinuous geological features (e.g., fault edges/tips, pinch-outs, fracture systems, channels, and small geobodies), where the values of the computed curvature become extremely high. However, while these high values can indicate the existence of non-reflecting objects, they do not deliver their specific geometric characteristics. In this study we present a novel method that better characterizes the shapes of these complex geological features by extending the assumption of local surfaces (2D surfaces in 3D space) into local hypersurfaces (3D hypersurfaces in 4D space), with their corresponding (three rather than two) principal (and effective) curvature parameters. We demonstrate the advantages of our method by comparing the conventional dip-based surface curvature parameters with the hypersurface curvature parameters, using a synthetic model/image with different types and shapes of geological features and two seismic images of real data containing a complex fault and hidden buried channels.

## INTRODUCTION

Seismic- and model-driven curvature parameters are effective attributes that provide geoscientists with enhanced images of complex structural features, such as different types and shapes of reflecting horizons (e.g., synclines/anticlines, ridges/troughs, and saddles), faults and even subtle fractures, and stratigraphic features, such as fine layering, channels, and mounds. Along with the other types of seismic/model-driven attributes, they reveal improved structural and stratigraphic aspects of subsurface reservoirs (e.g., Chopra and Marfurt, 2005, 2015; Hart and Sagan, 2007; Smith, 2008). Local volumetric subsurface curvatures are normally defined as functions depending on changes in the inline and crossline dips of surfaces, at fine three-dimensional grid points (in the seismic resolution), where the scalar field value of each grid point represents seismic-driven or model-driven reflectivities (or diffractivities). The local surfaces are parameterized as quadratic surfaces, and the computed curvatures quantify the degree to which the surfaces deviate from being planar (e.g., Chopra and Marfurt, 2007a). The volumetric structural curvature attributes measure the curvedness of the bending and folding of (real or virtual) seismic reflector surfaces (e.g., Verma et al., 2018). Furthermore, volumetric estimation of curvatures alleviates the need for picking horizons, which is especially useful in regions where no continuous surfaces exist (Chopra and Marfurt, 2007c; Chen et al., 2012).

For low to moderate structural complexities, curvature values are relatively small and can be reliably computed from the local surface assumption. On the other hand, in subsurface regions affected by strong folding and sharp faulting due to lateral tectonic stresses and vertical compaction, as well as areas affected by considerable weathering/erosion, the assumption of the existence of local surfaces is no longer valid, and the curvatures are abruptly raised to extreme values. Although these high-value curvatures can serve as indicators of sharp discontinuities (e.g., Lisle, 1994; Hart, 2002; Mai et al., 2009a; Li, 2014; Chopra et al., 2019, Liang, 2019; Wang et al., 2022, among others), they do not provide actual information about the characteristics (types and shapes) of the local objects in these areas.

In this study, we present a novel method that, in addition to the ability to better indicate sharp discontinuous geological objects, also delivers useful additional information about the characteristics of these complex geological features. This is achieved by extending the assumption of local surfaces (2D surfaces in 3D space) into local hypersurfaces (3D hypersurfaces in 4D space), with their corresponding (three rather than two) principal and effective curvature parameters.

Many papers have been published on methods for computing curvatures and their geophysical/geological interpretation. The classic book by Chopra and Marfurt (2007a) on seismic attributes, and the references within, provide extensive descriptions of the existing methods for computing curvatures with many worldwide case studies, demonstrating the high value of this type of seismic attribute for prospect identification and reservoir characterization.

High values of principle curvatures (major and minor) are commonly used to identify faults (e.g., Chopra and Marfurt, 2007a); however, they contain vast amounts of information that can sometimes be confusing since these values can be assigned to many different types of discontinuous objects (Roberts, 2001). As a remedy, to better see the lineaments contained within the target surfaces, the most positive and most negative curvatures were introduced (e.g., Hakami et al., 2004; Di and Gao, 2016a; Zhao et al.,2018; Gaci et al., 2021). Unlike the principal curvatures that depend on both the gradient and the Hessian of the of the prospective local surfaces, the most positive and most negative curvatures only depend on the Hessian (Roberts, 2001; Chopra and Marfurt, 2010b). Geologic structures can be studied at a variety of scales and generally exhibit curvatures of different wavelengths (Waldron and Snyder, 2020). In particular, the most positive and most negative curvatures can be both short-wavelength and long-wavelength (Chopra and Marfurt, 2010a, 2015; Hunt et al, 2018).

Obviously, in cases where the seismic horizons are already interpreted and explicitly given as $z = z(x, y)$, the computation of the curvature attributes along these surfaces is straightforward. However, the main challenge is to establish the curvature parameters directly from the available volumetric seismic (or its related attributes, e.g., the envelope) image field(s) prior to any interpretation process. Furthermore, computing curvatures at different scales can reveal geological features which should guide both structural and stratigraphic interpretation (Al-Dossary and Marfurt, 2006; Chopra and Marfurt, 2006a). Hence, several volumetric-driven methods have been proposed to compute the surface curvature attributes at each sample of the 3D seismic (or seismic attribute) grid. These methods normally assume the availability of the gradient vector of the volumetric scalar field at each point (representing a reflector-normal vector), or its orientation parameters, such as the local dip and azimuth angles or the inline and crossline dips, and, if necessary, also the gradient magnitude (e.g., Blumentritt et al., 2006).

Donias et al. (1998) proposed estimating the curvature by applying the divergence operator to the dip-azimuth vector field where the calculation is performed in the local normal planes. Stewart and Wynn (2000) studied the structural curvature spectrum (in the wavenumber domain) and concluded that a rapid decay of the curvature with the wavelength allows reliable curvature measurements of long-wavelength features even if short-wavelength elements (such as faults) or noise are present. Roberts (2001), Al-Dossary and Marfurt (2006), Chopra and Marfurt (2006b, 2007b), Mai and Marfurt (2008), Guo et al. (2010), and Mai et al. (2014), among others, extracted the parameters needed to compute the curvature at each sample of the 3D volume using the fractional derivatives of the apparent dip on each time slice, where the seismic parameter becomes a function of two variables. Sigismondi and Soldo (2003) used curvature attribute maps to visualize and interpret structural features in several case studies and demonstrated that a higher contrast in the curvature attributes helps to better identify the fault system. West et al. (2003) presented a method where individual curvatures are computed as horizontal gradients of the apparent dip for several directions and are then combined to generate the final curvature volume. However, the results obtained by these methods can be meaningful and reliable only if the dip estimates are good enough. Flierman et al. (2008), Chopra and Marfurt (2008, 2014, 2016), and Nissen et al. (2009) applied curvature attributes to highlight faults, fractures, flexures, and channels, based on the expectation that higher curvatures correlate with increased fracture density. Klein et al. (2008)

proposed a robust approach, where a local quadratic surface that best fits the data is analyzed in the proximity of each 3D grid point. The curvatures are computed by estimating the shapes of local quadratic reflecting surfaces associated with the 3D grid points within defined horizontal ranges, where the surface depths $z(x, y)$ correspond to the maximum cross-correlation value between the central trace and the surrounding traces. In a number of case studies, Mai et al. (2009b) applied volumetric curvatures and other geometric attributes to find lineaments corresponding to the edges of reef-related structures.

Fracture planes are often normal to a principal direction of the minor or major curvature. Generally, a curvature computed along a certain azimuth mainly highlights structures in the direction perpendicular to that azimuth (Boe and Daber, 2010). In this cited work, the curvature volumes were computed for different azimuths (60 degrees apart) and then co-rendered in an RGB (Red, Green, Blue) color map. These authors also showed that by combining azimuthal curvatures in a unique volume, the overall noise can be reduced (Daber and Boe, 2010). Case studies by Hunt et al. (2010) revealed statistically significant correlations of the AVAz (amplitude vs. azimuth), VVAz (velocity vs. azimuth), and curvature attributes with the image log fracture density mapped at the centimeter and millimeter scales. The local stresses from the image logs and their interrelationship with the curvature attributes were further explored by Yenugu and Marfurt (2011). Choi et al. (2011) studied the most positive and most negative dip-based curvatures of the scalar gravity field in the Dead Sea Basin to obtain detailed information of the crustal density structures in that region. Staples (2011) and Chopra and Marfurt (2011, 2012, 2013) applied structural, amplitude, and other types of curvatures to extract useful geological information. For the structural curvature, they used the seismic amplitude scalar field to compute the first derivatives of the inline and crossline components of the structural dip, while for the amplitude curvature, they computed the first derivatives of the gradient components of the energy-weighted amplitudes, representing directional measures of amplitude variability. Prezzi et al. (2012, 2016) studied the curvatures (along with the other geophysical attributes) of gravimetric, magnetic, and resistivity anomalies in Bajada del Diablo, Argentina, and concluded that Pampa Sastre Formation conglomerates could have been ejected and/or displaced during excavation by the impact of an extraterrestrial projectile. Guo et al. (2012) suggested a workflow to quantitatively correlate the fracture-related curvature lineaments from different azimuths to the production data. Martins et al. (2012) computed volumetric curvatures considering seismic horizons as level surfaces of an implicit function, viewed as a local surface identifier. Chen et al. (2012) suggested a new algorithm based on the instantaneous frequency and studied 3D multi-scale volumetric curvatures. Karimi and Fomel (2013) converted the seismic image to a special coordinate frame where horizons/reflector surfaces are defined explicitly at any point. Giroldi and Garossino (2014) applied a modified fractional derivative algorithm to compute long-wavelength curvature attributes, especially useful for the interpretation of structural and stratigraphic features. Rich and Marfurt (2013), Di and Gao (2014a), and Bhattacharya and Verma (2019) studied the third-order surface parameters in order to compute the gradients of the curvature and demonstrated that these parameters may improve surface and fault characterization. In the latter study, the results from the most positive and most negative curvature attributes revealed several types of aligned faults with different azimuthal orientations. Hale (2013) computed images of fault likelihoods, strikes, and dips from 3D seismic data, and then extracted fault surfaces (across which seismic events appeared discontinuous yet correlated), and estimated fault throws by a constraint minimization of sample values on the opposite sides of the fault. Valdmanis (2013) applied the dip-based curvature attributes for automatic highlighting of salt domes in seismic images (which are of particular interest in seismic exploration, as they are often linked to hydrocarbon finds). Gao (2013) applied the curvature of seismic reflectors to identify areas of enhanced potential to form tensile fractures where he used curvature gradients to identify areas of shear fractures. Holubnyak et al. (2014) applied curvature attributes to infer fracture swarms, fracture sets, flexures, sags, and paleokarst. In later studies, Gao and Di (2015) and Di and Gao (2016b, 2017) demonstrated that the maximum curvature and flexure (derivative of the curvature with respect to (wrt) the arclength) can be effective parameters to characterize the intensity and orientation of faults and fractures. Roden et al. (2015) studied several categories of seismic attributes, including instantaneous amplitude-accentuating inversion and geometric attributes (semblance, eigensystem-based coherency, similarity, different types of curvatures) and applied principal component analysis (PCA) to deduce which seismic attributes or combination of seismic attributes have meaningful interpretive significance. Hunt et al. (2011, 2018) used the curvature attributes, in particular, the most positive curvature (with certain parameterization and filtering), to predict the key properties of geologic targets that can potentially yield hazardous natural fractures. Silva et al. (2016) demonstrated that the azimuthally-dependent curvature of the seismic data can be understood as a second-order directional derivative and applied as an oriented edge-detection filter. Guo et al. (2016) studied the Barnett Shale of the Fort Worth Basin and found that in the survey acquired prior to hydraulic fracturing, AVAz anomalies were stronger and highly correlated with major structural lineaments measured by curvature attributes. Fun et al. (2017) computed the curvature attributes of different scales, varying the resolution of differential operators, thus revealing structural characteristics of different ranges, and integrated the results in a single RGB fusion. Hill et al. (2017) found statistically significant correlations between volumetric most positive curvature and natural fracture density indicated from high-resolution image log data. Qi and Marfurt (2018) analyzed volumetric aberrancy (which is the lateral gradient vector of the curvature, characterized with magnitude and azimuth) to delineate subtle faults and flexures of the Fort Worth Basin, Texas, whose throw falls below the seismic resolution. Di et al. (2018) and Di and AlRegib (2019) suggested a method to estimate reflector dips based on the waveform curvature and flexure analysis. Ashraf et al. (2020) applied neural networks and computed multiple structural attributes, including dip and curvature, filtered and unfiltered similarities, thinned fault likelihood, fracture density, and fracture proximity, in order to identify small-scale faults and their orientations. Cao et al. (2020) provided a method to evaluate seismic curvature attribute-based coal textures that makes it possible, along with other attributes, to distinguish between undeformed, cataclastic, and granulated coal. Liao et al. (2020) applied the most negative and most positive curvatures for detection of synclinal and anticlinal flexures, respectively, related to the existing reverse faults, which allowed (along with the use of other seismic attributes, like reflection amplitude, dip-azimuth, and seismic variance) to identify the subsurface structure, and characterize the composite damage zones of tight sandstone reservoirs in northeast Sichuan, China. Chopra and Marfurt (2020a, 2020b) applied inline and crossline multi-spectral dip components to compute the curvature parameters, which allowed them to define lineaments or channel/reef edges. Buck et al. (2007), Hameed and Hermana (2020), and Hussein et al. (2021) used multi-seismic attributes, and in particular, the curvature of seismic data to characterize fractured basement reservoirs with high production potential. Deshuang et al. (2021) calculated curvatures using seismic data to predict fractured reservoirs and then generated rose diagrams to predict fracture orientations. Li et al. (2021) studied fault openings of a shale oil reservoir using the information of lost circulation, fracturing, channeling, and formation dip angle during real drilling, and they found that open, semi-open, and sealed faults have their own characteristics in the most positive and most negative curvature attributes. Verma et al. (2022) applied amplitude-based attributes, in particular the amplitude curvature, to identify faults and sand channels that cannot be detected with structural curvature analysis. In a case study, Lin et al. (2022) analyzed the correlation between seismic attributes computed from pre- and post-stack seismic data with the interpreted azimuth obtained from image logs and micro-seismic data and demonstrated that the strike of the most positive curvature can be used (along with the azimuthal anisotropy) as an indicator for the azimuth of natural fractures within the study area. Nantanoi et al. (2022) explored the seismic data from the Bowland Shale formation, UK, applying three seismic attributes: similarity, spectral decomposition, and curvature to the seismic volume. They demonstrated that tracking the similarity and spectral decomposition attributes makes it possible to detect normal and thrust faults, where the decomposition may be especially useful at higher frequencies. However, the curvature attributes also display the offsets from faults and remnants of ancient channels, which do not show up on the similarity and spectral decomposition attributes. Ray et al. (2022) performed a case study in Nova Scotia, applied seismic data conditioning, and compared the structure and the amplitude curvature attributes. They demonstrated that the structural curvature alone clearly depicts the fault patterns but does not delineate the channels due to the absence of any flexure across the channel. Chopra et al. (2022) computed coherence, Euler curvature, aberrancy, and fault likelihood to delineate subtle basement fractures, and demonstrated that these attributes can also aid in constructing the low-frequency velocity model needed for impedance inversion.

Sun et al. (2022) applied volumetric aberrancy to provide geologically interpretable fracture delineation in fault-controlled Karstic reservoirs in Sichuan Basin, China. Detection of middle to small-scale fractures related to fault-controlled reservoirs is reliable with the volumetric aberrancy (Qi and Marfurt, 2018; Sun et al., 2022) that represents the lateral gradient vector of the structural curvature. The magnitude of the aberrancy characterizes the intensity of the curvature variation, while the azimuth defines the direction of its rapid change. The volumetric aberrancy can be computed for a variety of curvature attributes: hypersurface- and dip-based, most positive and most negative, principal, effective, curvature along the structural dip, etc. Shuai et al. (2023) introduced the curvature attributes to compute the horizontal strains, and then applied the elastic plate theory to estimate the corresponding stresses, thus taking into account the geological structure.

Machine learning (ML) has been recently widely applied for the interpretation of seismic attribute volumes, in particular, those related to reflection discontinuities: semblance, coherency, curvature, and variance, which can be used to detect and characterize faults (Roden et al., 2021). From a computer vision perspective, faults are a special group of edges, and deep learning (DL) techniques, mainly represented by convolutional neural networks (CNN) have been applied to edge detection (Zhao and Mukhopadhyay, 2018). Ao et al. (2021) proposed a deep learning-based volumetric curvature extraction approach that directly derives structural curvature volumes from the seismic image volume. However, obviously, the accuracy and reliability of ML/DL-driven attributes highly depend on the labeling and training of the chosen network with the corresponding geological features and characteristics. The proposed method for computing hypersurface curvatures can be an attractive tool for enhancing the fidelity of the ML/DL labeling/training stages.

In this paper, we suggest a different and more universal approach to computing the curvature parameters at 3D subsurface grid points, where the input scalar field at those points can represent different types and scales of subsurface geological parameters. In the seismic interpretation workflow, the commonly used 3D scalar fields are seismic amplitudes of migrated images or their attributes, such as their analytic trace attributes (e.g., envelope and instantaneous phase), or their inverted model parameters, such as seismic-driven impedance. The background interval velocity field used in seismic migrations to obtain the seismic image volumes can also be used to provide the long-wavelength orientation and curvature information.

The proposed method is formulated for any $n$-dimensional space as a hypersurface in a space of a higher dimension $n+1$, provided the normal line and the tangent hyperplane can be defined for any given point of a smooth hypersurface. The curvature can be defined for any "azimuth" in the tangent hyperplane. For $n=2$ (a 2D surface in 3D space), only two (the minimum and the maximum) principal curvatures exist; for $n=3$ (a 3D hypersurface in 4D space), there is an additional (the medium-valued) principal curvature. We show that the corresponding three effective curvature parameters: mean (average of the three principal curvatures), Gaussian (product of the three principal curvatures), and "scalar" (average of the three pair-wise products of the principal curvatures) are the most informative attributes, each illuminating different geological features, for example, making it possible to separate between "smooth" structural horizons and various discontinuous objects.

In this paper, we first define the curvature parameters for a regular 2D surface in 3D space and then extend this definition to a general case of any dimension, in particular, the case of a 3D hypersurface in 4D space. We introduce the matrices of the first and second fundamental forms of the hypersurface and provide the relationships for computing their components. Next, we formulate a specific generalized eigenvalue problem and reduce it to the standard eigensystem analysis using the square root matrix. Finally, we provide three examples: (1) A synthetic model containing structural and seismic-driven representations of different continuous and discontinuous geological objects, (2) A field example from New Zealand containing a complex fault and a system of buried channels, and (3) A field example from the North Sea with an extensive environment featuring a major strike-slip fault and several normal growth faults dipping in opposite directions.

The hypersurface-based approach to the volumetric curvature attributes was briefly introduced in the SEG and IAMG Abstracts (Koren et al., 2022; Tertois et al., 2022). In this paper, we provide a detailed mathematical derivation of the method, present two real-data examples, and suggest a structural interpretation of the computed results.

Derivations and auxiliary background material have been moved to the appendices.

Appendices A and B include the relevant background material for this study. In Appendix A we review the common curvature parameters used by geoscientist interpreters (listed, for example, in the work by Chopra and Marfurt 2007a). All of them, defined initially for an explicit regular 2D surface, are extended to volumetric curvatures in a straightforward way.

In Appendix B we explain the conventional method for computing the volumetric curvature parameters of a 3D scalar field without explicitly extracting horizons. We convert the dip-based formulae into relationships providing better numerical stability.

In Appendix C we convert the curvature, expressed initially as the generalized Rayleigh quotient, into the standard Rayleigh quotient, whose stationary values and directions follow from the eigensystem analysis.

In Appendix D we describe the computation of the square root for the inverse matrix of the first fundamental form of the hypersurface curvature ($\mathbf{A}^{-1/2}$), required for the conversion described in Appendix C.

In Appendix E we list the explicit formulae for the coefficients of the characteristic quadratic/cubic equations for the surface/hypersurface dimensions $n = 2$ and $n = 3$, respectively, whose roots are the principal curvatures. For $n = 3$, these coefficients are the effective hypersurface curvatures (mean, scalar, and Gaussian), and the relationships are especially useful in cases where only the principal curvatures are needed without their corresponding principal directions, or in order to verify the results of the eigensystem analysis.

In Appendix F we provide a numerical example for computing the principal hypersurface curvatures and their corresponding 3D azimuths and full 4D principal directions for a single point of a smooth 3D scalar field.

In Appendix G we explain the spherical angles for the 4D space geometry. The full 4D eigenvectors can be presented as points on a 3D sphere of a unit radius in 4D space. Unlike 3D space with two spherical angles, in 4D space, there are three spherical angles.

In Appendix H we explain the Fourier approach used in this study for smoothing (filtering) the scalar field volume and for accurately computing the gradient and the Hessian components of the scalar field, needed to establish the curvatures.

In Appendix I we compare the computational complexities of the dip-based and hypersurface principal curvatures, considering analytical and numerical approaches.

In Appendix J we demonstrate a direct analogy between the principal curvatures (and their directions) and the natural frequencies (and modes) of the free oscillations in mechanical and electrical systems.

# SURFACE AND HYPERSURFACE CURVATURES

In appendices A and B we review the conventional surface-based seismic curvatures commonly used in seismic interpretation (e.g., Chopra and Marfurt 2007a). In Appendix A the curvature parameters are defined for an explicit 2D surface $z(x, y)$, and in Appendix B the curvatures are extended to volumetric parameters at a given subsurface point using a scalar field defined on a 3D grid. The classic concept for computing surface curvatures is based on a normal plane which is defined as one of the planes containing the surface normal vector at a given 3D grid point. Thus, for a fixed point, the surface curvature depends on the azimuthal direction of the normal plane (Gauss, 2019), which is the curvature of the curved line resulting from the intersection between the local surface and the normal plane. Note that for smooth surfaces, there is a single tangent plane and an infinite number of normal planes with different azimuths. The azimuth can be measured either in the tangent plane (i.e., in the local reference frame whose "vertical" axis is the surface normal) or in the horizontal plane $xy$ (i.e., in the global reference frame), with respect to a reference azimuth given in the plane of the measurement. We refer to this azimuth as the normal plane azimuth, and we measure it in the global frame, where it has three components. However, it is sufficient to specify only two components in the horizontal plane, $\mathbf{r}_h = \begin{bmatrix} r_x & r_y \end{bmatrix}$ where $r_x = \cos\psi$, $r_y = \sin\psi$, and $\psi$ is the global azimuth angle. The vertical component $r_z$ of the tilted azimuth can then be restored (computed) considering that the tilted azimuth vector $\mathbf{r} = \begin{bmatrix} r_x & r_y & r_z \end{bmatrix}$ belongs to the tangent plane and is therefore perpendicular to the surface normal $\nabla F(x, y, z)$. This definition can be extended for 3D hypersurfaces in 4D space with four "azimuthal" components, where we can specify three components of the curvature eigenvectors in the horizontal hyperplane $\begin{Bmatrix} r_x & r_y & r_z \end{Bmatrix}$, and then compute the missing fourth "vertical" component from the orthogonality condition.

A hypersurface is an $n$-dimensional manifold (a topological variety) embedded in an $n+1$ dimensional space (e.g., Lee, 2009). This definition is also valid for regular 2D surfaces in 3D space and planar curves.

A scalar value **aMb**, where **M** is a matrix and **a**, **b** are vectors, is called a bilinear form; a scalar value **aMa**, where **M** is a *square* matrix, is called a quadratic form (e.g., Gantmacher, 2012). The azimuthally dependent curvature $k$ of a hypersurface can be presented as the generalized Rayleigh quotient, which is a ratio of two quadratic forms,

$$k(\mathbf{r}) = \frac{\mathbf{r}\,\mathbf{B}\,\mathbf{r}}{\mathbf{r}\,\mathbf{A}\,\mathbf{r}} \qquad . \qquad (1)$$

In this equation, **A** and **B** are the square matrices of the first and second fundamental forms of the curved surface, respectively (defined below), and **r** is the global azimuth vector of the curvature (e.g., Hartmann, 2003). These matrices are also referred to as the first and second fundamental tensors of the surface. The dimension $n$ of the two matrices **A** and **B** and vector **r** corresponds to that of the surface (or hypersurface) while $n+1$ is the space dimension; in other words, for 2D surfaces in 3D space $n=2$, and for 3D hypersurfaces in 4D space $n=3$.

Since both the numerator and the denominator in equation 1 are second-degree homogeneous polynomials wrt the components of vector **r**, the fraction is zero-degree homogeneous, and one may assume without any loss of generality that vector **r** is normalized, $\mathbf{r} \cdot \mathbf{r} = 1$. For a regular 2D surface in 3D space, its components are just the sine and cosine of the projection of the normal plane azimuth vector onto the horizontal plane.

When explicitly defining a 2D surface, $z = z(x, y)$ (an interface between two geological layers), or a 3D hypersurface, $v = v(x, y, z)$ (a scalar field representing, for example, seismic reflectivity, impedance, or even velocity), it is convenient to assume that the coordinates (the axes of the section or the volume) and the scalar field have the same units (normally the units of distance). For example, in cases where the scalar spatial field represents seismic velocity, we normalize the seismic velocity by a dominant frequency $f$ (with typical values ranging between 10 to 60 Hz), where the normalized scalar field, i.e., the ratio $v(x, y, z) / f$, represents a spatially varying wavelength, with the units of distance. Throughout the paper, we use different types of scalar fields, and the normalization factor $f$ is changed accordingly such that the units of the normalized scalar field are distance units.

Remark: The spatial coordinates $x, y, z$ are defined along axes of the "horizontal" hyperplane, and $v/f$ is the altitude axis of the 4D space: $(x, y, z, v)$. The hypersurface $v = v(x, y, z)$ splits the hyperspace (4D space) in two sub-spaces: $v < v(x, y, z)$ and $v > v(x, y, z)$.

In differential geometry, the matrix of the first fundamental form of a surface/hypersurface depends on the gradient components of the scalar field,

$$\mathbf{A} = \mathbf{I} + \frac{\nabla v \otimes \nabla v}{f^2} \qquad , \qquad (2)$$

where $\mathbf{I}$ is the identity matrix of dimension $n$, while the matrix of the second fundamental form includes the gradient length and Hessian components of the scalar field,

$$\mathbf{B} = \frac{\nabla\nabla v}{\sqrt{f^2 + \nabla v \cdot \nabla v}} \qquad . \qquad (3)$$

The gradient of the scalar field $\nabla v$ is a vector of length $n$, and the Hessian $\nabla\nabla v$ is a symmetric matrix of size $n$. Note that matrix $\mathbf{A}$ is symmetric and positive definite, while matrix $\mathbf{B}$ is only symmetric and may have real eigenvalues of any sign. Matrix $\mathbf{A}$ is the metric tensor of the curvilinear hypersurface; it defines the line and area elements, $ds$ and $dS$, respectively,

$$
\begin{aligned}
ds &= \sqrt{d\mathbf{x} \cdot \mathbf{A} \cdot d\mathbf{x}} \quad \text{where} \quad d\mathbf{x} = \begin{bmatrix} dx_1 & dx_2 & \cdots & dx_n \end{bmatrix} \quad , \\
dS &= \sqrt{\det \mathbf{A}}\, dx_1 dx_2 \ldots dx_n \quad \text{where} \quad \det \mathbf{A} = 1 + \frac{\nabla v \cdot \nabla v}{f^2} \quad .
\end{aligned}
\qquad (4)
$$

Equations $1 - 4$ are valid for any dimension $n \geq 1$, where for $n = 1$ they are reduced to the standard formula for the curvature of a line $v(x)$ (rather than that of a surface/hypersurface),

$$k(x) = \frac{f^2 v''(x)}{\left[ f^2 + v'^2(x) \right]^{3/2}} \quad , \quad v' = \frac{dv}{dx} \quad \text{and} \quad v'' = \frac{d^2 v}{dx^2} \qquad , \qquad (5)$$

and the area element reduces to an arclength element. Note that for a geometric line, the normalization factor is $f = 1$.

Our goal is to find the stationary values, $k_i$, $i = 1,\ldots,n$, of the curvature in equation 1, and their corresponding principal azimuthal directions $\mathbf{r}_i$. In Appendix C, we show that this can be achieved by converting the generalized Rayleigh quotient (equation 1) into the standard Rayleigh quotient, which reduces the task to a problem of eigenvalues and eigenvectors. The stationary values of the curvature in equation 1 are the eigenvalues of a symmetric (but not necessarily positive definite) matrix $\mathbf{C}$,

$$\mathbf{C} = \mathbf{A}^{-1/2}\,\mathbf{B}\,\mathbf{A}^{-1/2} \qquad , \qquad (6)$$

while the principal directions can be computed from the eigenvectors of this matrix.

## CHARACTERISTIC EQUATION FOR THE PRINCIPAL CURVATURES

The eigensystem analysis is suitable when we need both the curvatures and their principal directions, and when the hypersurface dimension $n$ is arbitrary. If only the eigenvalues are needed for $n = 2$ (a 2D surface in 3D space) or $n = 3$ (a 3D surface in 4D space), then the coefficients of the characteristic polynomial equation (quadratic or cubic, respectively) can be computed analytically. We applied the method suggested by Goldman (2005) and obtained the results listed in Appendix E.

## COMPUTING CURVATURES OF A HYPERSURFACE

In Appendix F we provide an example for a detailed, step-by-step, numerical computation of the hypersurface (principal) curvatures and their corresponding directions, for a single grid point of a scalar field volume. Since the principal directions are normalized to the unit length, it may be suitable to represent them in the form of polar angles in a 4D spherical frame of reference, as explained in Appendix G. The process of smoothing (filtering) the scalar field and computing its first and second derivatives is performed in the wavenumber domain and is explained in Appendix H.

In the next section, we explain the technique applied to model the geological features of the synthetic model used in the synthetic numerical example.

## MODELING STRUCTURE WITH GEOLOGICAL FEATURES

The synthetic model covers an area of 10 km by 7 km and ranges in depth from zero to 5 km. We select 14 surfaces representing realistic large and fine-scale horizon geometries with abrupt orientation changes. We arrange them in eight different stratigraphic sequences, with erosional and baselap relationships, to create discontinuities between the sequences with pinch-out geometries (Figure 1a). Four large, nearly vertical faults affect the western part of the model. In the eastern part, a salt body intrudes into the lower layers, and 18 faults radiate around the top of the salt body, affecting the horizons as the intrusion progresses towards shallower layers. The salt body and radial faults are courtesy of the RING Team and ASGA consortium.

Relative Geological Time

To create a 3D structural model, we first build a tetrahedral grid of the survey area, including the fault surfaces and the salt body. From the horizon data, we then compute the relative geological time (RGT) in the eight stratigraphic sequences (Mallet, 2004, 2014). The RGT is monotonous within each of the stratigraphic sequences and is discontinuous across faults and stratigraphic discontinuities (Figure 1b). From the RGT, we then compute the paleo-geographic coordinates $u$ and *v* which enable us to fully link the geological space (*x, y, z*) to the depositional space (*u, v, t*) where geological horizons are flat and unfaulted (Mallet, 2004, 2014).

<u>Velocity and Density</u>

We model physical rock properties in the depositional space (*u, v, t*), where the complexity linked to tectonic activity is removed, using straightforward spatial correlations. Since in this domain horizons are flat and continuous, we can model the velocity and density in a Cartesian grid representing the depositional space and correlate the properties with a constant variogram aligned with the grid axes.

We initialize the density in four locations close to the corners of the depositional space as a function of depth (Athy, 1930), increasing from $2.1 \times 10^3$ kg/m³ at the top to $2.6 \times 10^3$ kg/m³ at the bottom. We then perform a Sequential Gaussian Simulation (SGS) (Goovaerts, 1997; Chilès and Delfiner, 2012) using the pre-initialized density values as constraints, to simulate the subsurface heterogeneity and obtain a realistic increase in density with depth. Spatial correlations in the depositional space are governed by a 12 km × 12 km × 0.5 km Gaussian variogram, which ensures stronger lateral correlations within layers and large variability from one layer to the next in the vertical direction. Density values are simulated from a uniform distribution ranging from $2.1 \times 10^3$ kg/m³ to $2.8 \times 10^3$ kg/m³ (Figure 1c).

In one of the upper layers, we model a channel after an aerial photograph of the Amazon river, stacking channel and point bar deposits over a height of approximately 50 meters with slight lateral variations in geometry to better represent channel sections, i.e., narrower towards the bottom of the channel. Since the layers are flat in the depositional space, adding a channel is a straightforward painting process. In the areas labeled as channel deposits, we simulate density by SGS with a 10 km×10 km×0.3 km variogram and a Gaussian distribution with a mean value of $2.2 \times 10^3$ kg/m³ and a standard deviation of 50 kg/m³. The density of point bar deposits is simulated by SGS with the same variogram and a uniform density distribution ranging from $2.3 \times 10^3$ kg/m³ to $2.6 \times 10^3$ kg/m³. A depth slice in the resulting density volume at the depth level of the channel is shown in Figure 1d.

Using the UVT Transform which maps the depositional space to the geological space, we transport the simulated density to the geological space, thus adding the structural complexity of faulting and folding arising from tectonic activity. We also set density inside the salt body to a constant value of $2.2 \times 10^3$ kg/m³. Figure 2a shows the final density model in geological space. The density, which increases with depth, is strongly correlated along layers and varies rapidly in the direction normal to the layers. The channel facies create areas with sharp contrast. At this stage, we also add a few small spherical density anomalies (simulating "point" diffractors) in one of the upper layers, where just five voxels, arranged in a spherical shape, have a higher density than the background rock. These are not easily visible in the density volume.

We proceed in a similar way to model the velocity field in the depositional space, using a 15 km×15 km×0.1 km variogram to obtain a stronger lateral continuity and higher variations in the vertical direction. We group the layers according to their stratigraphic sequences and use the distributions set out in Table 1 to simulate velocities by SGS.

Channel facies' velocities are simulated with a 10 km×10 km×0.3 km variogram and a uniform distribution ranging from 3 km/s to 3.3 km/s. Point bar facies' velocities are simulated using the same variogram and a Gaussian distribution with a mean value of 3.5 km/s and a standard deviation of 125 m/s. Using the UVT Transform, we transport the flat velocity volume to the geological space and obtain the velocity model shown in Figure 2b.

Impedance and Seismic Amplitude

At each voxel location in geological space, we multiply the simulated density and velocity to obtain impedance values. As we know the position of the faults, salt body, and stratigraphic unconformities in the geological space from the structural model, we can smooth the impedance across these boundaries using a simple Gaussian convolution kernel to avoid creating overly sharp contrasts. We obtain the impedance volume shown in Figure 2c.

Finally, using the procedure described by Vizeu et al. (2022), we iterate over each vertical trace in the impedance volume and compute the reflectivity from the impedance. We then convolve the trace with a wavelet to generate seismic amplitudes and obtain the volume shown in Figure 2d.

## NUMERICAL EXAMPLES

Synthetic Model

The synthetic model described in the previous section is used in this example for computing the proposed curvatures. The survey grid includes 496 inlines and 348 crosslines, with a lateral resolution of 20 m, and 826 depth samples with a vertical resolution of 6 m. The original velocity of compressional waves varies in range $2.129 \text{ km/s} \leq v \leq 4.5 \text{ km/s}$.

Figures 3, 4, and 5 are devoted to the curvatures of the synthetic velocity volume. Figures 3a and 4a show the original unsmooth velocity, and Figures 3b and 4b show the smooth velocity. We first compute the velocity gradient and Hessian components, followed by the principal and effective

curvatures. The principal curvatures include the minor and major curvatures of the hypersurface and the 2D surface (dip-based curvatures) and an additional medium-valued curvature for the hypersurface. The effective curvatures include the mean and the Gaussian, and the additional scalar curvature for the hypersurface. In Figures 3c and 4c, we plot the mean curvature of the hypersurface, while in Figures 3d and 4d – the mean curvature of the 2D surface (dip-based curvature). In Figures 3e and 4e, we plot the Gaussian curvature of the hypersurface, while in Figures 3f and 4f – the Gaussian curvature of the 2D surface. The scalar curvature exists for 3D hypersurfaces only and has no analog for regular 2D surfaces; it is plotted in Figure 5.

Figures 6, 7, and 8 show the curvatures of the synthetic seismic envelope. The input data is considered the real part of the analytic signal (whose spectrum includes the non-negative frequencies only). We apply the Hilbert transform and compute first the matching imaginary part, and then the unsmooth and smooth envelopes. Figures 6a and 7a show the (real) input signal, while in Figures 6b and 7b we plot the smooth envelope. Next, we compute the first and second derivatives of the smooth envelope and its principal and effective curvatures. Figures 6c / 6e and 7c / 7e show the mean/Gaussian curvatures of the hypersurface, while Figures 6d / 6f and 7d / 7f – the mean/Gaussian curvatures of the 2D surface. In Figure 8 we show the scalar curvature of the hypersurface seismic envelope.

Comparison of Figures 3c / 4c with 3d / 4d for the velocity field and Figures 6c / 7c with 6d / 7d for the seismic envelope shows that the mean curvature of the hypersurface is much more continuous along geological structures than the dip-based mean curvature. Both, fine variations inside layers and stratigraphic discontinuities, are highlighted with a high resolution. The boundary of the salt body is completely recovered, even in parts where its dip is close to the dip of the background stratigraphic layers. The dip-based mean curvature is noisier than the hypersurface mean curvature, less continuous, fails to highlight the whole boundary of the salt body, and does not enhance the channel enough to individualize it as a geological object.

The Gaussian curvature of the hypersurface shown for the velocity field in Figures 3e / 4e and for the seismic envelope in Figures 6e / 7e highlights discontinuities in the input. Compared to the dip-based Gaussian curvature in Figures 3f / 4f / 6f / 7f, discontinuities of the hypersurface curvature, such as faults and stratigraphic unconformities, are sharper and less noisy, while the stratigraphic structure is dimmed further. The channel feature above the stratigraphic unconformity in the top third of the model stands out as a set of bright spots where the cross-section intersects the object (red arrows).

The point diffractors, added as small, dense spheres in the density volume, appear in the mean and Gaussian hypersurface curvatures of the seismic envelope as bright spots on a clean background (green arrows), whereas these objects appear on the level of noise in the dip-based curvatures.

The scalar hypersurface curvature shown in Figures 5 and 8 combines characteristics from the mean and Gaussian curvatures, retaining a strong continuity along stratigraphic structures and highlighting discontinuities.

Seismic Image of the Parihaka dataset, New Zealand

The Parihaka dataset, provided by the New Zealand Crown Minerals and available on dataunderground.org, consists of good quality seismic acquired offshore New Zealand, in the Taranaki Basin, an area which provides most of New Zealand's oil and gas production. This Cretaceous basin contains mostly marine sediments and many channels filled with younger terrestrial sediment.

The grid of the Parihaka seismic image includes 640 inlines with a resolution of 25 m, 1480 crosslines with a resolution of 12.5 m, and 401 depth samples with a resolution of 5 m (up to the maximum depth of 2 km). In this example, we computed and plotted the curvatures of the seismic amplitudes of the image traces rather than the envelope.

In Figure 9 we show two vertical sections: inline and crossline, selected from the seismic amplitude volume, and in Figure 10 we show a horizontal slice, where both figures include the region of interest. Figures 9a / 10a show the original seismic image (the input). Figures 9b / 10b and 9c / 10c show the cross-sections of the hypersurface and the dip-based mean-curvature volumes, respectively. Figures 9d / 10d and 9e / 10e show the cross-sections of the hypersurface and the dip-based Gaussian curvatures, respectively. Figures 11a and 11b show the vertical sections and the horizontal slice of the hypersurface scalar curvature, respectively.

In the 3D views (Figures 9b / 9c, inline and crossline sections), the hypersurface mean curvature shows better continuity along reflectors than the dip-based curvature, highlighting horizons and transition zones. As demonstrated in Appendix B, the dip-based curvature effectively ignores the second vertical derivative of a scalar field, $\partial^2 v / \partial z^2$, which is not optimal for horizon detection. In the slice views (Figures 10b / 10c), the hypersurface mean curvature demonstrates better continuity as well, and it highlights channel edges and a relay area between two faults. In both vertical (Figures 9d / 9e) and horizontal (Figures 10d / 10e) sections, the Gaussian curvature of the hypersurface shows better definition of the faults and is less noisy than the dip-based Gaussian curvature. The hypersurface scalar curvature (Figures 11a / 11b) clearly highlights faults and edges of channels with a good definition.

Figure 12 shows the seismic amplitude co-rendered with the absolute values of three effective hypersurface curvatures or two effective dip-based curvatures, each mapped to a different RGB channel. The two upper plots are related to the hypersurface curvatures. The 3D view of the vertical inline and crossline sections is to the left (Figure 12a) and the horizontal slice view is to the right (Figure 12b). The mean curvature, mapped to the blue channel, highlights the geological structures which are mainly discontinuous in the vertical direction (hence, more easily seen on vertical sections). The Gaussian curvature, mapped to the green channel, mostly picks discontinuities such as faults and edges of channels (where all three principle curvatures are significant). The scalar curvature, mapped to the red channel, shows a mix of structure and discontinuous characteristics. Spots (e.g., point-like diffractors), where the three effective curvatures have high values, stand out in white and represent volumetric features where there are discontinuities with no preferred structure orientation (e.g., the spots inside the fault zones). Between the two faults, the mean curvature shows the deformed area affected by fault growth that retains some dipping geological structure. The two lower plots are related to the dip-based effective curvatures. The 3D view is to the left (Figure 12c) and the horizontal slice view is to the right (Figure 12d). The mean curvature is mapped to the blue channel, and the Gaussian curvature – to the green. The dip-based surfaces have no scalar curvature, so the red color is absent in these plots. The hypersurface effective curvatures have a better definition than the corresponding dip-based curvatures, and therefore, the RGB plot for the effective hypersurface curvatures is much sharper than that for the dip-based surfaces. Furthermore, mapping the additional scalar hypersurface curvature to the red channel enhances areas where all three effective curvatures are high by making them white. We can see that adding a third hypersurface-based attribute on the red channel helps visualizing these crucial features.

A channel and a fault are shown separately in Figure 13. Figures 13a and 13c show the mean curvature of the hypersurface and the dip-based mean curvature, respectively, in the proximity of the channel; Figure 13b shows the corresponding seismic amplitude. The hypersurface mean curvature shows better continuity even in the low energy: The internal structure of the large channel, which cannot be seen in the input signal (Figure 13b), is clearly visible with the hypersurface mean curvature (Figure 13a). This is not so for the dip-based curvature (Figure 13c). Figures 13d and 13f show the mean curvature of the hypersurface and the dip-based mean curvature, respectively, in the proximity of a fault. Again, the hypersurface mean curvature shows better continuity even in the low energy: The internal structure of the fault area, which is blurred in the input signal (Figure 13e), is clearly visible with the hypersurface mean curvature (Figure 13d), which is not the case with the dip-based curvature (Figure 13f).

Figure 14 shows a joint view of the vertical and horizontal cross-sections of the mean curvature volume. Figures 14a and 14b are related to the hypersurface and the dip-based curvatures, respectively. Channel features can be traced on the depth (horizontal) slice and correlated with the internal structure visible on the vertical slice, which is highlighted much more strongly and clearly by the hypersurface mean curvature than by the dip-based mean curvature. The faulted area in the right part of the figure also correlates well between the horizontal slice and the vertical cross section, and deformed and ruptured stratigraphic layers in that area are more strongly detailed by the hypersurface mean curvature.

<u>Seismic Image of the Clyde dataset, North Sea</u>

The Clyde field in the North Sea is located in an extensive environment featuring a major strike-slip fault and several normal growth faults dipping in opposite directions on either side of this main fault (Gibbs, 1984). The seismic data used in this study is a good-quality subset comprising 281 crosslines, 791 inlines and 388 depth samples (Figure 15a). Figures 15b – 15g display different curvatures on an inline section across the strike-slip fault zone. Figure 15b shows the input seismic amplitude used to compute the curvature attributes. Figures 15d and 15f show this seismic amplitude overlain with the mean and Gaussian curvatures computed with the standard dip-based method, respectively. Figures 15c, 15e, and 15g show the results of our new hypersurface-based curvature formulation. The mean curvature is very different in the dip-based (15d) and hypersurface-based (15e) methods. Because it represents the curvature of the full 3D amplitude attribute instead of deriving from dip and azimuth information only, the hypersurface mean curvature shows extremely strong continuity along geological structures. This is especially visible in areas where structures were deformed by the strike-slip fault (inside the white box). The hypersurface Gaussian curvature (15g) has a better definition and more precisely highlights discontinuities than the dip-based Gaussian curvature (15f). Finally, the third effective curvature produced by the hypersurface method, the scalar curvature, both highlights continuous structures and enhances their termination in discontinuous areas.

Figure 16 shows how a blending of curvature attributes can provide insights to interpreters. In the left-hand column, absolute values of standard dip-based Gaussian and mean curvatures are mapped to the green and blue color channels, respectively. In the right-hand column, absolute values of hypersurface scalar, Gaussian and mean curvatures are mapped to the red, green and blue color channels, respectively. Figures 16a and 16b show a crossline section through one of the normal growth faults, Figures 16c and 16d show an inline section through the strike-slip fault area, and Figures 16e and 16f show a depth slice featuring both the strike-slip and normal faults. Compared to the standard dip-based curvatures, the hypersurface curvatures provide more information with the addition of a third color channel. The mean curvature mapped to the blue channel captures horizons and geological layers. Flexure points (white arrows in Figures 16a – 16d) are clearly visible. Discontinuities produce high values of the Gaussian curvature and stand out in green shades. Areas where continuous structures terminate are characterized by high values of the three hypersurface curvatures and are highlighted in light, almost white colors. On the depth slice (Figure 3f), the normal faults (white rectangles) appear more continuous on the hypersurface display. Inside the strike-slip fault area, the hypersurface curvatures provide more detail and more continuity in enhanced geological features (white arrows).

## DISCUSSION

Interpreters already have a wide range of attributes derived from seismic data to help them characterize geological features. Our proposed hypersurface-based curvatures rely on a different approach: they are computed from any scalar field and their formulation is equivalent in all three directions. Our governing equations include vectors, tensors, and their invariants, rather than their components. These factors make them suitable to better detect and quantify geological objects which are not naturally characterized as smooth surfaces or planar faults. In all of our examples, the hypersurface mean curvature highlights geological features that strongly contrast with their immediate surroundings. Stratigraphic layers with strong impedance contrast to the previous and following layers are shown as extrema of hypersurface mean curvature with strong continuity in the lateral directions, even where the seismic signal has low energy, such as in fault zones. Channels are detected as elongated connected areas of either high or low hypersurface mean curvature. By contrast, the hypersurface Gaussian curvature shows strong extrema where geological features are discontinuous, such as point diffractors, fault planes/volumes, and contacts between channel sediments and the surrounding rock. Finally, the hypersurface scalar curvature shows a mix of structure and discontinuous characteristics, and in addition, it reveals intermediate geological features which are not characterized by the other two curvatures (e.g., rough surfaces, faults, and channels).

## CONCLUSIONS

We suggest a novel approach to compute volumetric curvatures, where the 3D volume is presented as a hypersurface in 4D space. The fourth dimension (the altitude axis of the hyperspace) is the studied scalar parameter (converted to the units of distance), which may be a seismic reflectivity, its envelope and unwrapped phase, wave velocity, impedance, relative geological time (RGT), or any other scalar field. Unlike conventional volumetric curvature methods (surface-driven methods based on the derivatives of the inline and crossline dips), where the lateral and vertical coordinates are treated differently, our governing equations for the principal (and effective) curvatures treat the partial derivatives of the scalar field parameter with respect to all the Cartesian coordinates in a symmetric way: The hypersurface curvature formulae, based on the invariants of the field gradient and Hessian, are coordinate-free. With the suggested approach, the vertical (altitude) direction has no special preference; the computed curvatures are insensitive to a 3D volume rotation. The input data are the nodal values of the scalar field defined on a fine regular 3D grid. This leads to a simple, transparent, and straightforward algorithm to compute the three effective hypersurface curvatures (mean, scalar, and Gaussian), the three stationary (principal) hypersurface curvatures (minor, medium-valued, and major), and their corresponding principal directions at each grid point. This, in turn, makes it possible to characterize (quantify) non-surface-like geological features, such as fault edges and tips, fractures, small geobodies, and point-like diffractors. The advantages of the hypersurface-driven curvatures are demonstrated along a synthetic model containing different continuous and discontinuous geological objects and along real seismic image volumes containing channel and fault features. The suggested approach provides high-resolution contrast images with detailed and rich information for interpreters. Unlike the classic dip-based method, the hypersurface-driven method provides meaningful curvatures at points with vanishing gradients (but not necessarily vanishing Hessians), where the inline and crossline dips do not exist or are not clearly defined.

## ACKNOWLEDGMENT

The authors are grateful to Aspen Technology for the financial and technical support of this study, and for the permission to publish its results. Gratitude is extended to our colleagues Lorena Guerra and Beth Orshalimy, and to the reviewers and editors of GJI for valuable remarks and comments that helped to improve the content and the style of this paper.

## DATA AVAILABILITY

The Parihaka input data used in this study belongs to the Government of New Zealand, is an open source, and can be reached using this link (https://dataunderground.org/dataset/parihaka). The Clyde input data is not publicly available. The output data of both sets is intellectual property of Aspen Technology, this data is confidential and cannot be shared due to legal and commercial reasons.

# APPENDIX A. STANDARD CURVATURE PARAMETERS OF A 2D SURFACE

In this appendix, we review the most commonly used curvature parameters (e.g., listed in Chopra and Marfurt, 2007a, chapter 4, pages 73-75; they are also explained by Rich, 2008). The parameters are initially defined for an explicit 2D surface $z(x, y)$, and then extended to volumetric parameters in a straightforward way.

Consider a local quadratic surface $z(x, y)$ in the proximity of a central (reference) point $x = x_\text{o}, y = y_\text{o}$, defined explicitly with the five coefficients $(a, b, c, d, e)$,

$$z(x_\text{o} + \Delta x, y_\text{o} + \Delta y) = z(x_\text{o}, y_\text{o}) + a(\Delta x)^2 + b(\Delta y)^2 + c\,\Delta x\,\Delta y + d\,\Delta x + e\,\Delta y \qquad , \qquad \text{(A1)}$$

where $\Delta x$ and $\Delta y$ are small lateral intervals. The derivatives at the central point are,

$$\frac{\partial z}{\partial x} = d \equiv p \quad , \qquad \frac{\partial^2 z}{\partial x^2} = 2a = \frac{\partial p}{\partial x} \quad , \qquad \frac{\partial^2 z}{\partial x \partial y} = c = \frac{\partial p}{\partial y} = \frac{\partial q}{\partial x} = \frac{1}{2}\left(\frac{\partial p}{\partial y} + \frac{\partial q}{\partial x}\right) \qquad , \qquad \text{(A2)}$$

$$\frac{\partial z}{\partial y} = e \equiv q \quad , \qquad \frac{\partial^2 z}{\partial y^2} = 2b = \frac{\partial q}{\partial y} \quad ,$$

where the surface's first derivatives, $d$ and $e$, can be associated with the volumetric inline and crossline dips, $p$ and $q$, that can be computed from the gradient components of the 3D scalar field. Hence, the substitutions in equation A2 make it possible to convert the surface curvature parameters listed below into the volumetric parameters, explained in Appendix B.

The **mean** and **Gaussian** curvatures can be obtained from the matrices of the first and second fundamental forms (matrices **A** and **B** in equations 2 and 3). In the case of a regular 2D surface, the normalization factor in equations 2 and 3 is set to one, $f = 1$, and these matrices read,

$$\mathbf{A} = \mathbf{I} + \nabla z \otimes \nabla z = \begin{bmatrix} 1+d^2 & d\,e \\ d\,e & 1+e^2 \end{bmatrix} \quad , \quad \mathbf{B} = \frac{\nabla \nabla z}{\sqrt{1+\nabla z \cdot \nabla z}} = \frac{1}{\sqrt{1+d^2+e^2}} \begin{bmatrix} 2a & c \\ c & 2b \end{bmatrix} \quad . \quad \text{(A3)}$$

Next, the square root of the inverse of matrix $\mathbf{A}$ is computed,

$$\mathbf{A}^{-1/2} = \begin{bmatrix} \dfrac{d^2+e^2\sqrt{1+d^2+e^2}}{\left(d^2+e^2\right)\sqrt{1+d^2+e^2}} & \dfrac{1-\sqrt{1+d^2+e^2}}{\left(d^2+e^2\right)\sqrt{1+d^2+e^2}}\,d\,e \\[4mm] \dfrac{1-\sqrt{1+d^2+e^2}}{\left(d^2+e^2\right)\sqrt{1+d^2+e^2}}\,d\,e & \dfrac{e^2+d^2\sqrt{1+d^2+e^2}}{\left(d^2+e^2\right)\sqrt{1+d^2+e^2}} \end{bmatrix} \quad , \quad \text{(A4)}$$

to obtain the resulting symmetric matrix (tensor), $\mathbf{C} = \mathbf{A}^{-1/2}\,\mathbf{B}\,\mathbf{A}^{-1/2}$ (equation 6). The effective curvatures are the invariants of tensor $\mathbf{C}$. The **mean** curvature is one-half of its trace, while the **Gaussian** curvature is its determinant,

$$k_{\text{mean}} = \frac{\text{tr}\,\mathbf{C}}{2} = \frac{a\left(1+e^2\right)+b\left(1+d^2\right)-c\,d\,e}{\left(1+d^2+e^2\right)^{3/2}} \quad , \quad k_{\text{Gauss}} = \det\mathbf{C} = \frac{4a\,b-c^2}{\left(1+d^2+e^2\right)^2} \quad . \quad \text{(A5)}$$

The two **principal** curvatures are the eigenvalues of matrix (tensor) $\mathbf{C}$,

$$k_{\text{major}} = k_{\text{mean}} + \sqrt{k_{\text{mean}}^2 - k_{\text{Gauss}}^2} \quad , \quad k_{\text{minor}} = k_{\text{mean}} - \sqrt{k_{\text{mean}}^2 - k_{\text{Gauss}}^2} \quad . \quad \text{(A6)}$$

The **maximum** and **minimum** curvatures are defined as,

$$k_{\max} = \begin{cases} k_{\text{major}} \ \text{if } \left|k_{\text{major}}\right| \geq \left|k_{\text{minor}}\right| \\ k_{\text{minor}} \ \text{if } \left|k_{\text{major}}\right| < \left|k_{\text{minor}}\right| \end{cases} \quad , \quad k_{\min} = \begin{cases} k_{\text{minor}} \ \text{if } \left|k_{\text{major}}\right| \geq \left|k_{\text{minor}}\right| \\ k_{\text{major}} \ \text{if } \left|k_{\text{major}}\right| < \left|k_{\text{minor}}\right| \end{cases} \quad . \quad \text{(A7)}$$

Note that the maximum and minimum curvatures are defined wrt their *absolute* values: Following the definitions of equations A6 and A7, in case the mean curvature is positive, the maximum curvature is the major and the minimum is the minor; otherwise, (in case it is negative), the maximum curvature is the minor and the minimum is the major. The **most extreme** curvature is either the maximum or the minimum curvature, depending on which absolute value prevails (Di and Gao, 2014b).

The **most positive** and **most negative** curvatures are defined as,

$$k_{\text{pos}} = (a+b) + \sqrt{(a-b)^2 + c^2} \quad , \quad k_{\text{neg}} = (a+b) - \sqrt{(a-b)^2 + c^2} \quad , \quad \text{(A8)}$$

both can be positive or negative,

$$
\begin{aligned}
\left\{ a < 0, \quad b < 0, \quad c^2 < 4ab \right\} \quad &\rightarrow \quad \left\{ k_{\text{pos}} < 0, \quad k_{\text{neg}} < 0 \right\} , \\
\left\{ a > 0, \quad b > 0, \quad c^2 < 4ab \right\} \quad &\rightarrow \quad \left\{ k_{\text{pos}} > 0, \quad k_{\text{neg}} > 0 \right\} , \\
\left\{ ab > 0 , \quad\quad\quad c^2 > 4ab \right\} \quad &\rightarrow \quad \left\{ k_{\text{pos}} > 0, \quad k_{\text{neg}} < 0 \right\} , \\
ab < 0 \quad &\rightarrow \quad \left\{ k_{\text{pos}} > 0, \quad k_{\text{neg}} < 0 \right\} .
\end{aligned}
\quad \text{(A9)}
$$

Unlike the principal curvatures (equation A6) which are the eigenvalues of tensor **C** (that includes both first and second derivatives), the most positive and most negative curvatures are the eigenvalues of the Hessian matrix of the explicit 2D surface defined in equation A1 (including only second derivatives),

$$\nabla \nabla z(x, y) = \begin{bmatrix} \dfrac{\partial^2 z}{\partial x^2} & \dfrac{\partial^2 z}{\partial x \partial y} \\ \dfrac{\partial^2 z}{\partial x \partial y} & \dfrac{\partial^2 z}{\partial y^2} \end{bmatrix}_{x=y=0} = \begin{bmatrix} 2a & c \\ c & 2b \end{bmatrix} \quad , \quad \text{(A10)}$$

These complementary curvature parameters have recently attracted geoscientists as useful attributes for structural interpretation; the simultaneous analysis of their signs provides important insight regarding the type of the investigated local 2D surface (Table 2, Chopra and Marfurt, 2007a).

To establish the **dip** and **strike** curvatures, the corresponding azimuthal directions,

$$\psi_{\text{dip}} = +\arctan\frac{e}{d} \quad , \qquad \mathbf{r}_{\text{dip}} = \begin{bmatrix} \cos\psi_{\text{dip}} \\ \sin\psi_{\text{dip}} \end{bmatrix} = \frac{1}{\sqrt{d^2+e^2}}\begin{bmatrix} +d \\ +e \end{bmatrix} \quad ,$$

$$\psi_{\text{strike}} = -\arctan\frac{d}{e} \quad , \qquad \mathbf{r}_{\text{strike}} = \begin{bmatrix} \cos\psi_{\text{strike}} \\ \sin\psi_{\text{strike}} \end{bmatrix} = \frac{1}{\sqrt{d^2+e^2}}\begin{bmatrix} -e \\ +d \end{bmatrix} \quad . \tag{A11}$$

are computed. The use of equation 1 yields,

$$k_{\text{dip}} = \frac{\mathbf{r}_{\text{dip}} \cdot \mathbf{B} \cdot \mathbf{r}_{\text{dip}}}{\mathbf{r}_{\text{dip}} \cdot \mathbf{A} \cdot \mathbf{r}_{\text{dip}}} \quad , \qquad k_{\text{strike}} = \frac{\mathbf{r}_{\text{strike}} \cdot \mathbf{B} \cdot \mathbf{r}_{\text{strike}}}{\mathbf{r}_{\text{strike}} \cdot \mathbf{A} \cdot \mathbf{r}_{\text{strike}}} \quad , \tag{A12}$$

where **A** and **B** are the above-mentioned (equation A3) matrices of the first and second fundamental forms of the surface, leading to the explicit formulae for the dip and strike curvatures,

$$k_{\text{dip}} = \frac{2\left(a\,d^2 + b\,e^2 + c\,d\,e\right)}{\left(d^2+e^2\right)\left(1+d^2+e^2\right)^{3/2}} \quad , \qquad k_{\text{strike}} = \frac{2\left(a\,e^2 + b\,d^2 - c\,d\,e\right)}{\left(d^2+e^2\right)\left(1+d^2+e^2\right)^{1/2}} \quad . \tag{A13}$$

Koenderink and van Doorn (1992) introduced two new curvature characteristics that may be applied as alternatives or additions to the effective and/or principal curvatures of a 2D surface: the curvedness and the shape index.

The **curvedness** is a positive number, with the units of reciprocal distance, that characterizes only the amount of the curvature, but not the shape of the curved surface. This parameter represents the root-mean-square (RMS) value of the two principal curvatures,

$$c_k = \sqrt{\frac{k_{\text{major}}^2 + k_{\text{minor}}^2}{2}} \qquad . \qquad (A14)$$

Unlike the mean and the Gaussian curvatures, the curvedness vanishes only at locally planar points of curved surfaces (where both effective and both principal curvatures vanish).

On the contrary, the **shape index** is a scale-invariant parameter (for example, all spheres with different radii have the same index -1; the index is independent of the units of length),

$$s = \frac{2}{\pi} \arctan \frac{k_{\text{minor}} + k_{\text{major}}}{k_{\text{minor}} - k_{\text{major}}} \quad , \quad k_{\text{major}} \geq k_{\text{minor}} \quad , \quad -1 \leq s \leq +1 \qquad . \qquad (A15)$$

Note that the denominator in equation A15 is negative. For a planar surface, the shape index does not exist. According to the cited study, two shapes for which the shape index differs merely by the sign, represent complementary pairs that will fit together as 'stamp' and 'mold' when suitably scaled.

The **azimuths of the principal curvatures** (measured in the horizontal plane from axis $x$ counterclockwise) are computed by considering the vanishing derivative of the curvature in equation 1 wrt the azimuth,

$$\psi_{1,2} = \pm \frac{1}{2} \arccos \frac{m}{\sqrt{m_c^2 + m_s^2}} + \frac{1}{2} \arctan \left( m_s, m_c \right) \qquad , \qquad (A16)$$

where,

$$\begin{aligned} m_c &= -2(a+b)\,d\,e + c\left(2 + d^2 + e^2\right) , \\ m_s &= -2a\left(1 + e^2\right) + 2b\left(1 + d^2\right) , \\ m &= 2(a-b)\,d\,e - c\left(d^2 - e^2\right) . \end{aligned} \qquad (A17)$$

The inverse tangent of the two arguments in equation A16 is defined as,

$$\text{if} \quad \beta = \arctan\left(m_s, m_c\right) \quad, \quad \text{then} \quad \begin{aligned} \cos\beta &= \frac{m_c}{\sqrt{m_c^2 + m_s^2}} \\ \sin\beta &= \frac{m_s}{\sqrt{m_c^2 + m_s^2}} \end{aligned} \quad . \tag{A18}$$

It is not easy to predict analytically which of the two principal azimuths in equation A16 is related to the minor curvature, and which to the major (the expression for the second derivative of the curvature wrt the azimuth is complicated). Hence, one must first compute numerically the curvature for both azimuths using equation 1, where $\mathbf{r} = \begin{bmatrix} \cos\psi & \sin\psi \end{bmatrix}$. Note that the two principal directions of equation A16 are not normal to each other in the horizontal plane. The principal directions, $\mathbf{v}_{1,2}\left(\psi\right)$,

$$\mathbf{v}\left(\psi_i\right) = \begin{bmatrix} \cos\psi_i & \sin\psi_i & z\left(\psi_i\right) \end{bmatrix} \quad, \quad \text{where} \quad z\left(\psi_i\right) = d\cos\psi_i + e\sin\psi_i \quad, \quad i = 1, 2 \quad, \tag{A19}$$

are only mutually orthogonal in the tangent plane where they also have a vertical component $z\left(\psi_i\right)$, and this component can be computed considering that vectors $\mathbf{v}_{1,2}$ belong to the tangent plane.

The principal directions in equation A19 are not normalized. As mentioned, in the tangent plane, these vectors are orthogonal,

$$\mathbf{v}_1 \cdot \mathbf{v}_2 = \mathbf{v}\left(\psi_1\right) \cdot \mathbf{v}\left(\psi_2\right) = \cos\left(\psi_1 - \psi_2\right) + \left(d\cos\psi_1 + e\sin\psi_1\right)\left(d\cos\psi_2 + e\sin\psi_2\right) = 0 \quad, \tag{A20}$$

where the principal azimuths in the horizontal plane, $\psi_1$ and $\psi_2$, are given in equations A16 and A17. The cross product $\mathbf{n} = \mathbf{v}_1 \times \mathbf{v}_2$ is the normal to the curved surface and to its tangent plane.

Note that in a principal reference frame $\begin{bmatrix} \mathbf{v}_1 & \mathbf{v}_2 & \mathbf{n} \end{bmatrix}$, the deviation of a 2D curved surface from the tangent plane simplifies to,

$$z\left(x_o + \Delta x, y_o + \Delta y\right) - z\left(x_o, y_o\right) = \frac{k_1\left(\Delta x\right)^2 + k_2\left(\Delta y\right)^2}{2} \qquad , \qquad \text{(A21)}$$

which can be considered an analog of equation A1. For a 3D hypersurface, the similar deviation reads,

$$\frac{v\left(x_o + \Delta x, y_o + \Delta y, z_o + \Delta z\right) - v\left(x_o, y_o, z_o\right)}{f} = \frac{k_1\left(\Delta x\right)^2 + k_2\left(\Delta y\right)^2 + k_3\left(\Delta z\right)^2}{2} \qquad , \qquad \text{(A22)}$$

where $k_i$ are the principal curvatures.

The **unsphericity** curvature describes the local deviation of a given surface (or hypersurface) from spherical (Shary,1995); it represents half the difference between the principal curvatures,

$$k_{unsph} = \frac{k_{major} - k_{minor}}{2} \qquad . \qquad \text{(A23)}$$

An additional volumetric parameter may be the **curvature of a gradient line**. For this, equation 5 can be applied, where the first and second derivatives are computed along the gradient direction,

$$v' = \sqrt{\nabla v \cdot \nabla v} \quad , \qquad v'' = \frac{\nabla v \cdot \nabla \nabla v \cdot \nabla v}{\nabla v \cdot \nabla v} \qquad . \qquad \text{(A24)}$$

For a vanishing gradient, this parameter does not exist.

The **Euler curvature** depends on the azimuth $\psi$ (Chopra and Marfurt, 2014),

$$k\left(\psi\right) = k_{pos} \cos^2\left(\psi - \psi_{pos}\right) + k_{neg} \sin^2\left(\psi - \psi_{pos}\right) , \quad \psi_{pos} - \psi_{neg} = \pi / 2 \qquad , \qquad \text{(A25)}$$

where $k_{pos}, k_{neg}$ are the most positive and most negative curvatures, $\psi_{pos}, \psi_{neg}$ are their corresponding mutually orthogonal azimuths; all azimuths in this equation are measured in the tangent plane.

As a curvature is the derivative of the normal direction angle wrt the arclength, in a similar way, a **flexure** is the derivative of the curvature wrt the arclength. For dip-based surfaces, there are two stationary (principal) curvatures, but three stationary flexures which repeat with the azimuthal period of $\pi$ with the opposite signs (Di and Gao, 2017).

## APPENDIX B. FROM SURFACE TO VOLUMETRIC CURVATURES: CONVENTIONAL APPROACH

Existing methods for computing volumetric curvature parameters consider that each grid point under investigation/analysis is in the vicinity of a local (real or virtual) 2D reflecting surface; note that these volumetric methods do not require explicit picking of local horizon surfaces (e.g., Smith, 2008). The classical volumetric approach assumes that the (image/model-domain) scalar field in a proximity of any fixed point, $v(\mathbf{x}) = v(x, y, z)$, can be implicitly represented as $v(\mathbf{x}) = v\big[z(x, y)\big]$ where $z = z(x, y)$ describes a local horizon surface. This assumption leads to a chain rule for the derivatives of the field wrt the lateral coordinates,

$$v = v\big[z(x, y)\big] \quad , \qquad \frac{\partial v}{\partial x} = \frac{\partial v}{\partial z} \cdot \frac{\partial z}{\partial x} \quad , \qquad \frac{\partial v}{\partial y} = \frac{\partial v}{\partial z} \cdot \frac{\partial z}{\partial y} \qquad , \qquad \text{(B1)}$$

which yields the lateral derivatives (slopes) of the horizon, related to the gradient components of the 3D scalar field, referred to as the inline and crossline dips, $p$ and $q$,

$$\frac{\partial z}{\partial x} = \frac{\partial v / \partial x}{\partial v / \partial z} \equiv p \quad , \qquad \frac{\partial z}{\partial y} = \frac{\partial v / \partial y}{\partial v / \partial z} \equiv q \qquad . \qquad \text{(B2)}$$

The volumetric curvatures of the local scalar field are (effectively) the curvatures of this implicit horizon surface and can be expressed in terms of the inline and crossline dips and their (first) derivatives, as explained and shown below.

Both seismic-based and model-based scalar fields can be used to compute the curvatures. To establish the volumetric curvature, the continuous scalar parameter, $v(\mathbf{x})$, is first smoothed,

$$v_\sigma(\mathbf{x}) = v(\mathbf{x}) * k_\sigma(\mathbf{x}) \quad , \qquad k_\sigma = \left(\sqrt{2\pi}\,\sigma\right)^{-n} \exp\left(-\frac{\mathbf{x} \cdot \mathbf{x}}{2\sigma^2}\right) \qquad , \qquad \text{(B3)}$$

where $k_\sigma$ is a Gaussian kernel and $\sigma$ is the standard deviation (the size of the smoothing window) which makes the filter insensitive to noise and irrelevant at scales smaller than the size of the window. The symbol * is the convolution operator, and $n = 2,3$ is the space dimension. A gradient structure tensor (GST), $\mathbf{J}$, is then computed (e.g., Weickert, 1998, 1999; Fehmers and Hocker, 2003; Wu and Janson, 2017),

$$\mathbf{J}_\rho = \mathbf{J}_\sigma * k_\rho \quad , \qquad \mathbf{J}_\sigma = w \nabla v_\sigma \otimes \nabla v_\sigma \qquad . \qquad \text{(B4)}$$

Parameter $\rho$ represents the characteristic window, over which the orientation of the normal to the hypothesized horizon surface is to be analyzed. The convolution symbols in equations B3 and B4 do not necessarily mean applying the corresponding numerical operator, as it may be more convenient to map both the data (scalar $v$ in equation B3, or components of the gradient tensor $\mathbf{J}_\sigma$ in equation B4) and the Gaussian kernels to the wavenumber domain and to perform a multiplication there, with the subsequent return map of the result back to the data space.

Parameter $w$ (introduced before smoothing with window $\rho$ ) is the weight factor suggested by Luo et al. (2006) for seismic amplitudes to overcome the instabilities associated with the gradient direction. It is equal to the squared power $P$ of the analytic signal (or its envelope $E$ raised to a power four),

$$w = P^2 = E^4 \quad , \quad E = \sqrt{\text{Re}^2 + \text{Im}^2} \qquad \qquad , \qquad (B5)$$

where Re and Im are the real and imaginary parts of the complex analytic signal. The real part is the input data, and the imaginary part is its 1D Hilbert transform, so that the complex value represents an analytic signal, with non-negative frequencies or wavenumbers alone (e.g., Taner et al., 1979). The 1D Hilbert transform is performed in the vertical dimension only, trace by trace, in the vertical time or depth domains.

The convolution of the structure tensor in equation B4 is performed for this tensor component-wise and represents a linear isotropic smoothing. Better smoothing control, with higher quality and robustness, can be achieved using nonlinear anisotropic smoothing (e.g., Weickert, 1999; Hale, 2011). With nonlinear isotropic smoothing, one can avoid smoothing in the proximity of the edges. Further applying nonlinear anisotropic smoothing, which is based on the corresponding diffusion equation, makes it possible to detect not only the location of the edges but also their orientation. Close to edges, anisotropic smoothing is performed along the edges but not across them (e.g., Deschamps et al., 2004). Considering the coupling between the GST components, an additional improvement may be obtained by smoothing them simultaneously, with a set of coupled partial differential equations (e.g., Dascal et al., 2007; Rosman et al., 2011).

After the smoothing is complete, the eigenvectors of the structure tensor $\mathbf{J}_\rho$ are computed, where the largest eigenvalue provides an estimate of the local normal vector, $\mathbf{s}$ (e.g., Marfurt and Rich, 2010; Chopra and Marfurt, 2020b). The normal vector defines the (unitless) inline and crossline dips (these ratios are not angles),

$$p = s_x / s_z \quad , \quad q = s_y / s_z \qquad , \qquad (B6)$$

which agrees with equation B2, while the inverse relationships are,

$$s_x = \frac{p}{\sqrt{1 + p^2 + q^2}} \;\;,\;\; s_y = \frac{q}{\sqrt{1 + p^2 + q^2}} \;\;,\;\; s_z = \frac{1}{\sqrt{1 + p^2 + q^2}} \qquad . \qquad (B7)$$

The curvatures include combinations of the dip components and their derivatives (Roberts, 2001; Al-Dossary and Marfurt, 2006). The **mean** and **Gaussian** curvatures are,

$$k_{\text{mean}} = \frac{\dfrac{1}{2}\dfrac{\partial p}{\partial x}\left(1 + q^2\right) + \dfrac{1}{2}\dfrac{\partial q}{\partial y}\left(1 + p^2\right) - \dfrac{1}{2}\left(\dfrac{\partial p}{\partial y} + \dfrac{\partial q}{\partial x}\right)pq}{\left(1 + p^2 + q^2\right)^{3/2}} \quad ,$$

$$k_{\text{Gauss}} = \frac{\dfrac{\partial p}{\partial x}\dfrac{\partial q}{\partial y} - \dfrac{1}{4}\left(\dfrac{\partial p}{\partial y} + \dfrac{\partial q}{\partial x}\right)^2}{\left(1 + p^2 + q^2\right)^2} \quad . \tag{B8}$$

The **minor** and **major principal** curvatures are given in equation A6.

The **most negative** and **most positive** curvatures are,

$$k_{\text{neg}} = \frac{1}{2}\left(\frac{\partial p}{\partial x} + \frac{\partial q}{\partial y}\right) - \frac{1}{2}\sqrt{\left(\frac{\partial p}{\partial x} - \frac{\partial q}{\partial y}\right)^2 + \left(\frac{\partial p}{\partial y} + \frac{\partial q}{\partial x}\right)^2} \quad ,$$

$$k_{\text{pos}} = \frac{1}{2}\left(\frac{\partial p}{\partial x} + \frac{\partial q}{\partial y}\right) + \frac{1}{2}\sqrt{\left(\frac{\partial p}{\partial x} - \frac{\partial q}{\partial y}\right)^2 + \left(\frac{\partial p}{\partial y} + \frac{\partial q}{\partial x}\right)^2} \quad . \tag{B9}$$

In the case of a vanishing gradient $\nabla v$ of the scalar field, tensor **A** of the first fundamental form becomes the identity matrix **I**, while tensor **B** of the second fundamental form reduces to the Hessian matrix (scaled by the constant factor $f$ in the denominator), as follows from equations 2 and 3. In this case, equation 6 reduces to $\mathbf{C} = \mathbf{B}$, and the principal curvatures reduce to the (scaled) minor and major eigenvalues of the Hessian. The curvatures of explicit 2D surfaces with a vanishing gradient were analyzed by Young and Evans (1978); they are called the most positive and most negative curvatures (Appendix A). Obviously, for 3D hypersurfaces, the Hessian matrix yields an additional medium-valued curvature. The most positive and most negative curvatures are often computed despite the fact that in a general case, the actual gradient is normally varying and generally does not vanish. As noted by Roberts (2001), unlike the principal and effective curvatures (involving both, the gradient and the Hessian) that contain a great deal of information and can sometimes be confusing, the most positive and most negative curvatures are simpler and clearer. Indeed, the surface's gradient in a given point vanishes in the tilted reference frame whose horizontal coordinate plane $xy$ coincides with the tangent plane, and the altitude axis $z$ is normal to the curved surface. Although the Hessian components in the tilted frame are different, its eigenvalues remain invariant. Moreover, there are some differences between the concepts of the "vanishing gradients" for surfaces and hypersurfaces. In the case of hypersurface curvatures, we deal with the vanishing gradient of a scalar field, $\nabla v$, while in the case of dip-based curvatures, we deal with the vanishing gradient of a surface, $\nabla z(x, y)$. This, in turn, means zero inline and crossline dips, $p$ and $q$ (in the tilted frame), while the lateral derivatives of the dips, $\partial p/\partial x, \partial p/\partial y, \partial q/\partial x$ and $\partial q/\partial y$, are taken into account for the curvature computation.

Generally, the most positive curvatures show anticlinal structures and up-thrown fault blocks, while the most negative curvatures indicate the synclinal and down-thrown fault blocks (Chopra and Marfurt, 2007c). The most positive and most negative curvatures characterize the extreme bending of the structure (Li et al., 2021).

The surface curvatures are azimuthally dependent. The strike azimuth is the azimuth of a horizontal line lying on the surface's tilted tangent plane. The dip is the steepest angle of descent on a tilted plane, measured between the normal to this plane and the vertical axis. The azimuth of the dip differs by 90° from that of the strike. According to equation B7, the dip angle $\theta_{\text{dip}}$ and the dip azimuth $\psi_{\text{dip}}$ read,

$$\theta_{\text{dip}} = \arccos \frac{1}{\sqrt{1 + p^2 + q^2}} \quad , \quad \psi_{\text{dip}} = \arctan(q, p) \quad , \quad \psi_{\text{strike}} = \psi_{\text{dip}} \pm \frac{\pi}{2} \quad , \quad \text{(B10)}$$

and the **dip- and strike-azimuth curvatures** are,

$$k_{\text{dip}} = \frac{\frac{\partial p}{\partial x} p^2 + \frac{\partial q}{\partial y} q^2 + \left( \frac{\partial p}{\partial y} + \frac{\partial q}{\partial x} \right) pq}{\left( p^2 + q^2 \right) \left( 1 + p^2 + q^2 \right)^{3/2}} \quad ,$$

$$k_{\text{strike}} = \frac{\frac{\partial p}{\partial x} q^2 + \frac{\partial q}{\partial y} p^2 - \left( \frac{\partial p}{\partial y} + \frac{\partial q}{\partial x} \right) pq}{\left( p^2 + q^2 \right) \sqrt{1 + p^2 + q^2}} \quad .$$

$$\text{(B11)}$$

The **azimuths of the principal curvatures**, measured from axis $x$, are given by equation A16, where

$$m_c = -\left( \frac{\partial p}{\partial x} + \frac{\partial q}{\partial y} \right) p\, q + \frac{1}{2} \left( \frac{\partial p}{\partial y} + \frac{\partial q}{\partial x} \right) \left( 2 + p^2 + q^2 \right) \quad ,$$

$$m_s = -\frac{\partial p}{\partial x} \left( 1 + q^2 \right) + \frac{\partial q}{\partial y} \left( 1 + p^2 \right) \quad ,$$

$$m = +\left( \frac{\partial p}{\partial x} - \frac{\partial q}{\partial y} \right) p\, q - \frac{1}{2} \left( \frac{\partial p}{\partial y} + \frac{\partial q}{\partial x} \right) \left( p^2 - q^2 \right) \quad .$$

$$\text{(B12)}$$

Note that contrast between the eigenvalues of the structure tensor may be used as a measure of coherence (Hale, 2009; Chopra and Marfurt, 2010b; Wu, 2017).

The direct computation of the inline and crossline dip components is unstable (may be subject to noise), unless some special measures, such as plane wave destructor (Fomel, 2002, 2007), are undertaken. The spatial derivatives of the dips are even more unstable, as computing numerical derivatives is an "ill" operation in itself. In this study, we suggest computing the dip-based curvatures, where we bypass the use of unstable dips. For this, in equation set B6, we multiply both the numerator and the denominator by the absolute value of the gradient. This leads to equation B2 for the two dip components, $p$ and $q$, expressed in terms of the field gradient vector.

The derivatives of the dip components wrt the lateral coordinates become,

$$\frac{\partial p}{\partial x} = \frac{\dfrac{\partial^2 v}{\partial x^2}\dfrac{\partial v}{\partial z} - \dfrac{\partial^2 v}{\partial x \partial z}\dfrac{\partial v}{\partial x}}{\left(\partial v / \partial z\right)^2} \quad , \quad \frac{\partial p}{\partial y} = \frac{\dfrac{\partial^2 v}{\partial x \partial y}\dfrac{\partial v}{\partial z} - \dfrac{\partial^2 v}{\partial y \partial z}\dfrac{\partial v}{\partial x}}{\left(\partial v / \partial z\right)^2} \quad ,$$

$$\frac{\partial q}{\partial x} = \frac{\dfrac{\partial^2 v}{\partial x \partial y}\dfrac{\partial v}{\partial z} - \dfrac{\partial^2 v}{\partial x \partial z}\dfrac{\partial v}{\partial y}}{\left(\partial v / \partial z\right)^2} \quad , \quad \frac{\partial q}{\partial y} = \frac{\dfrac{\partial^2 v}{\partial y^2}\dfrac{\partial v}{\partial z} - \dfrac{\partial^2 v}{\partial y \partial z}\dfrac{\partial v}{\partial y}}{\left(\partial v / \partial z\right)^2} \quad ,$$

(B13)

where $v(x, y, z)$ is the analyzed scalar field. Introduction of equation sets B2 and B13 into equation set B8 yields the **mean** and the **Gaussian** curvatures,

$$k_{\text{mean}} = \frac{A \, \text{sign}\left(\partial v / \partial z\right)}{2\left(\nabla v \cdot \nabla v\right)^{3/2}} \quad , \quad k_{\text{Gauss}} = \frac{B}{4\left(\nabla v \cdot \nabla v\right)^2} \quad , \quad (B14)$$

where,

$$A = \left(\frac{\partial v}{\partial x}\right)^2 \frac{\partial^2 v}{\partial y^2} + \left(\frac{\partial v}{\partial y}\right)^2 \frac{\partial^2 v}{\partial x^2} - 2\frac{\partial v}{\partial x}\frac{\partial v}{\partial y}\frac{\partial^2 v}{\partial x \partial y} + \left(\frac{\partial v}{\partial z}\right)^2 \left(\frac{\partial^2 v}{\partial x^2} + \frac{\partial^2 v}{\partial y^2}\right)$$
$$- \frac{\partial v}{\partial z}\left(\frac{\partial v}{\partial x}\frac{\partial^2 v}{\partial x \partial z} + \frac{\partial v}{\partial y}\frac{\partial^2 v}{\partial y \partial z}\right) \quad ,$$

(B15)

$$B = \frac{\partial v}{\partial x}\frac{\partial^2 v}{\partial y \partial z}\left(\frac{\partial v}{\partial y}\frac{\partial^2 v}{\partial x \partial z} - \frac{\partial v}{\partial x}\frac{\partial^2 v}{\partial y \partial z}\right) + \frac{\partial v}{\partial y}\frac{\partial^2 v}{\partial x \partial z}\left(\frac{\partial v}{\partial x}\frac{\partial^2 v}{\partial y \partial z} - \frac{\partial v}{\partial y}\frac{\partial^2 v}{\partial x \partial z}\right)$$

$$+ 4\frac{\partial v}{\partial x}\frac{\partial v}{\partial z}\left(\frac{\partial^2 v}{\partial x \partial y}\frac{\partial^2 v}{\partial y \partial z} - \frac{\partial^2 v}{\partial y^2}\frac{\partial^2 v}{\partial x \partial z}\right) + 4\frac{\partial v}{\partial y}\frac{\partial v}{\partial z}\left(\frac{\partial^2 v}{\partial x \partial y}\frac{\partial^2 v}{\partial x \partial z} - \frac{\partial^2 v}{\partial x^2}\frac{\partial^2 v}{\partial y \partial z}\right) \qquad \text{(B16)}$$

$$+ 4\left(\frac{\partial v}{\partial z}\right)^2 \left[\frac{\partial^2 v}{\partial x^2}\frac{\partial^2 v}{\partial y^2} - \left(\frac{\partial^2 v}{\partial x \partial y}\right)^2\right] \quad ,$$

and,

$$\nabla v \cdot \nabla v = \left(\frac{\partial v}{\partial x}\right)^2 + \left(\frac{\partial v}{\partial y}\right)^2 + \left(\frac{\partial v}{\partial z}\right)^2 \qquad . \qquad \text{(B17)}$$

The other curvatures (most positive/negative, etc.) can be obtained in a similar fashion.

The most negative/positive curvatures are,

$$k_{\text{neg/pos}} = \frac{C_o \mp \sqrt{C_1 + C_2 + C_3 + C_4}}{2\left(\partial v / \partial z\right)^2} \qquad , \qquad \text{(B18)}$$

where the upper sign corresponds to the most negative curvature, and,

$$C_o = -\left(\frac{\partial v}{\partial x}\frac{\partial^2 v}{\partial x \partial z} + \frac{\partial v}{\partial y}\frac{\partial^2 v}{\partial y \partial z}\right) + \frac{\partial v}{\partial z}\left(\frac{\partial^2 v}{\partial x^2} + \frac{\partial^2 v}{\partial y^2}\right) \quad ,$$

$$C_1 = -2\frac{\partial v}{\partial z}\left(\frac{\partial^2 v}{\partial x^2} - \frac{\partial^2 v}{\partial y^2}\right)\left(\frac{\partial v}{\partial x}\frac{\partial^2 v}{\partial x \partial z} - \frac{\partial v}{\partial y}\frac{\partial^2 v}{\partial y \partial z}\right) \quad ,$$

$$C_2 = -4\frac{\partial v}{\partial z}\frac{\partial^2 v}{\partial x \partial y}\left(\frac{\partial v}{\partial x}\frac{\partial^2 v}{\partial y \partial z} + \frac{\partial v}{\partial y}\frac{\partial^2 v}{\partial x \partial z}\right) \quad , \qquad \text{(B19)}$$

$$C_3 = +\left[\left(\frac{\partial v}{\partial x}\right)^2 + \left(\frac{\partial v}{\partial y}\right)^2\right] \cdot \left[\left(\frac{\partial^2 v}{\partial x \partial z}\right)^2 + \left(\frac{\partial^2 v}{\partial y \partial z}\right)^2\right] \quad ,$$

$$C_4 = \left(\frac{\partial v}{\partial z}\right)^2 \left[\left(\frac{\partial^2 v}{\partial x^2} - \frac{\partial^2 v}{\partial y^2}\right)^2 + 4\left(\frac{\partial^2 v}{\partial x \partial y}\right)^2\right] \quad .$$

The **dip angle and its azimuth** simplify to,

$$\theta_{\text{dip}} = \arccos \frac{\partial v / \partial z}{\sqrt{\nabla v \cdot \nabla v}} \quad , \quad \psi_{\text{dip}} = \arctan \left( \frac{\partial v}{\partial y}, \frac{\partial v}{\partial x} \right) \qquad . \tag{B20}$$

The **dip- and strike-azimuth curvatures** are,

$$k_{\text{dip}} = \frac{\left( \partial v / \partial z \right)^4 D}{2 \left[ \left( \partial v / \partial x \right)^2 + \left( \partial v / \partial y \right)^2 \right] \left( \nabla v \cdot \nabla v \right)^3} \quad ,$$

$$k_{\text{strike}} = \frac{\left( \partial v / \partial z \right) E}{2 \left[ \left( \partial v / \partial x \right)^2 + \left( \partial v / \partial y \right)^2 \right] \nabla v \cdot \nabla v} \quad , \tag{B21}$$

where,

$$D = \frac{\partial v}{\partial z} \left[ \left( \frac{\partial v}{\partial x} \right)^2 \frac{\partial^2 v}{\partial x^2} + \left( \frac{\partial v}{\partial y} \right)^2 \frac{\partial^2 v}{\partial y^2} + 2 \frac{\partial v}{\partial x} \frac{\partial v}{\partial y} \frac{\partial^2 v}{\partial x \partial y} \right]$$

$$- \left[ \left( \frac{\partial v}{\partial x} \right)^2 + \left( \frac{\partial v}{\partial y} \right)^2 \right] \left( \frac{\partial v}{\partial x} \frac{\partial^2 v}{\partial x \partial z} + \frac{\partial v}{\partial y} \frac{\partial^2 v}{\partial y \partial z} \right) \quad , \tag{B22}$$

$$E = \left( \frac{\partial v}{\partial x} \right)^2 \frac{\partial^2 v}{\partial y^2} + \left( \frac{\partial v}{\partial y} \right)^2 \frac{\partial^2 v}{\partial x^2} - 2 \frac{\partial v}{\partial x} \frac{\partial v}{\partial y} \frac{\partial^2 v}{\partial x \partial y} \qquad .$$

The **azimuths of the principal curvatures** are given by equation A16 where,

$$m_c = -\left[\left(\frac{\partial v}{\partial x}\right)^2 - \left(\frac{\partial v}{\partial y}\right)^2\right]\left(\frac{\partial v}{\partial x}\frac{\partial^2 v}{\partial y \partial z} - \frac{\partial v}{\partial y}\frac{\partial^2 v}{\partial x \partial z}\right) + 2\left[\left(\frac{\partial v}{\partial x}\right)^2 + \left(\frac{\partial v}{\partial y}\right)^2\right]\frac{\partial v}{\partial z}\frac{\partial^2 v}{\partial x \partial y}$$

$$-2\frac{\partial v}{\partial x}\frac{\partial v}{\partial y}\frac{\partial v}{\partial z}\left(\frac{\partial^2 v}{\partial x^2} + \frac{\partial^2 v}{\partial y^2}\right) - 2\left(\frac{\partial v}{\partial x}\frac{\partial^2 v}{\partial y \partial z} + \frac{\partial v}{\partial y}\frac{\partial^2 v}{\partial x \partial z}\right)\left(\frac{\partial v}{\partial z}\right)^2 + 4\frac{\partial^2 v}{\partial x \partial y}\left(\frac{\partial v}{\partial z}\right)^3 ,$$

$$m_s = -2\left[\left(\frac{\partial v}{\partial x}\right)^2 + \left(\frac{\partial v}{\partial z}\right)^2\right]\left(\frac{\partial v}{\partial y}\frac{\partial^2 v}{\partial y \partial z} - \frac{\partial v}{\partial z}\frac{\partial^2 v}{\partial y^2}\right) + 2\left[\left(\frac{\partial v}{\partial y}\right)^2 + \left(\frac{\partial v}{\partial z}\right)^2\right]\left(\frac{\partial v}{\partial x}\frac{\partial^2 v}{\partial x \partial z} - \frac{\partial v}{\partial z}\frac{\partial^2 v}{\partial x^2}\right) ,$$

$$m = +\left[\left(\frac{\partial v}{\partial x}\right)^2 + \left(\frac{\partial v}{\partial y}\right)^2\right]\left(\frac{\partial v}{\partial x}\frac{\partial^2 v}{\partial y \partial z} - \frac{\partial v}{\partial y}\frac{\partial^2 v}{\partial x \partial z}\right) - 2\frac{\partial^2 v}{\partial x \partial y}\left[\left(\frac{\partial v}{\partial x}\right)^2 - \left(\frac{\partial v}{\partial y}\right)^2\right]\frac{\partial v}{\partial z}$$

$$+2\left(\frac{\partial^2 v}{\partial x^2} - \frac{\partial^2 v}{\partial y^2}\right)\frac{\partial v}{\partial x}\frac{\partial v}{\partial y}\frac{\partial v}{\partial z} .$$

(B23)

Thus, the dip-based volumetric curvatures are expressed in terms of the gradient and Hessian components alone. To compute the dip-based curvatures, we need all gradient components and five Hessian components of the scalar field $v(x, y, z)$ (all of them, except $\partial^2 v / \partial z^2$), although the classical approach suggests establishing the inline and crossline dips and their first lateral derivatives, instead of explicitly computing the Hessian. The suggested hypersurface approach requires all gradient and Hessian components to compute the curvatures.

Note that equations B14 – B23 for the dip-based curvatures are symmetric wrt the lateral coordinates $x, y$ only. On the contrary, our suggested equations for the hypersurface curvatures are symmetric wrt all three Cartesian coordinates.

Comment: It follows from equation B14 that the dip-based curvatures do not exist for a region with a vanishing gradient (e.g., the seismic velocity in a salt body). Equation B6 imposes even stricter constraints: the vertical component of the gradient should not vanish, otherwise, the inline and crossline dips will not exist, along with the dip-based curvatures. In our numerical examples, we set these non-existing curvatures to zero. Unlike the dip-based curvatures, the hypersurface curvatures exist at the nodes with the vanishing gradient and may be zero or non-zero, depending on the Hessian at this node. In the case of the vanishing gradient, the matrix of the first fundamental form becomes the identity matrix, and that of the second fundamental form reduces to the scaled Hessian matrix. The hypersurface principal curvatures become the eigenvalues of the scaled Hessian (in other words, the major and minor principal curvatures reduce to the most positive and most negative curvatures, respectively). This is an additional advantage of the hypersurface approach: The vanishing gradient is a legitimate case for the governing formulae, and it does not necessarily lead to non-existing or vanishing curvatures. Of course, if the gradient vanishes within some continuous region rather than at an isolated point, then the Hessian will vanish within that region as well.

### APPENDIX C. STANDARD AND GENERALIZED RAYLEIGH QUOTIENTS

Using, for example, the square root method, the generalized Rayleigh quotient (equation 1) can be converted to the standard (classical) Rayleigh quotient. Starting with equation 1 and noting that the square root matrix $\mathbf{A}^{1/2}$, likewise matrix $\mathbf{A}$, is also symmetric (and positive definite), we can rearrange the azimuthally dependent curvature,

$$k(\mathbf{r}) = \frac{\mathbf{r}\,\mathbf{B}\,\mathbf{r}}{\mathbf{r}\,\mathbf{A}\,\mathbf{r}} = \frac{\mathbf{r}\,\mathbf{B}\,\mathbf{r}}{\mathbf{r}\,\mathbf{A}^{1/2}\mathbf{A}^{1/2}\,\mathbf{r}} \qquad . \qquad (C1)$$

Next, we introduce vector $\mathbf{z}$, where the two-way relations between $\mathbf{r}$ and $\mathbf{z}$ are,

$$\mathbf{z} = \mathbf{A}^{1/2}\,\mathbf{r} \quad , \qquad \mathbf{r} = \mathbf{A}^{-1/2}\,\mathbf{z} \qquad , \qquad (C2)$$

and the curvature becomes,

$$k(\mathbf{r}) = \frac{\mathbf{r}\,\mathbf{B}\,\mathbf{r}}{\mathbf{r}\,\mathbf{A}\,\mathbf{r}} = \frac{\mathbf{z}\,\mathbf{A}^{-1/2}\,\mathbf{B}\,\mathbf{A}^{-1/2}\,\mathbf{z}}{\mathbf{z}\cdot\mathbf{z}} = \frac{\mathbf{z}\,\mathbf{C}\,\mathbf{z}}{\mathbf{z}\cdot\mathbf{z}} \qquad , \qquad (C3)$$

where $\mathbf{C}$ is a symmetric matrix (with real eigenvalues),

$$\mathbf{C} = \mathbf{A}^{-1/2}\,\mathbf{B}\,\mathbf{A}^{-1/2} \qquad . \qquad (C4)$$

Raising matrix $\mathbf{A}$ to power $-1/2$ is explained in Appendix D; this can be done analytically.

Note that Cholesky decomposition (Benoit, 1924) is an alternative to the proposed square root method.

The right-hand side of equation C3, $k = \dfrac{\mathbf{z}\,\mathbf{C}\,\mathbf{z}}{\mathbf{z}\cdot\mathbf{z}}$, is the standard Rayleigh quotient independent of the magnitude of vector $\mathbf{z}$. Therefore, it can be considered a unit-length normalized vector, and the principal curvature search can be rearranged as a constrained optimization problem,

$$k(\mathbf{z}) = \mathbf{z}\,\mathbf{C}\,\mathbf{z} \quad , \qquad \mathbf{z}\cdot\mathbf{z} = 1 \qquad . \qquad (C5)$$

Both $\mathbf{r}$ and $\mathbf{z}$ are unitless vectors, where their components are the direction cosines in the original global space and in the transformed space, respectively, to be further referred to as space $\mathbf{r}$ and space $\mathbf{z}$.

The constrained optimization problem formulated in equation C5 can be solved by the Lagrangian multipliers method,

$$g(\mathbf{z}) = \frac{1}{2}\mathbf{z}\,\mathbf{C}\,\mathbf{z} - \frac{\lambda}{2}(\mathbf{z}\cdot\mathbf{z} - 1) \;\rightarrow\; \text{stationary} \qquad , \qquad (C6)$$

where the factor $1/2$ has been introduced for convenience, and the vanishing gradient of the scalar augmented target function $g(\mathbf{z})$ leads to,

$$\mathbf{C}\mathbf{z} = \lambda\mathbf{z} \ , \quad \mathbf{z}\cdot\mathbf{z} = 1 \qquad , \qquad \text{(C7)}$$

which is a standard eigenvalue problem. The principal curvatures are the eigenvalues $\lambda_i$ of matrix (tensor) $\mathbf{C}$, and the effective curvatures (mean, Gaussian, etc.) are the tensor invariants. The eigenvectors $\mathbf{w}_i$ of matrix $\mathbf{C}$ are the principal directions in space $\mathbf{z}$. In order to convert the eigenvectors to space $\mathbf{r}$, we apply the second equation of set C2 and obtain eigenvectors $\mathbf{u}_i$,

$$\mathbf{u}_i = \mathbf{A}^{-1/2}\mathbf{w}_i \qquad . \qquad \text{(C8)}$$

Note that vectors $\mathbf{w}_i$ (in space $\mathbf{z}$) are orthogonal to each other, as these are eigenvectors of the symmetric matrix $\mathbf{C}$. However, this is not so for their corresponding vectors $\mathbf{u}_i$ (in the original space $\mathbf{r}$) computed with equation C8, $\mathbf{u}_i^{(3)}\cdot\mathbf{u}_j^{(3)} \neq 0$, where the superscript shows the dimension of the vector (in this case, of 3D hypersurfaces). These scalar products only vanish in 4D spaces. The fourth component of the eigenvectors can be computed considering that all principal directions are in the tangent hyperplane and thus perpendicular to the hypersurface normal $\mathbf{n}$,

$$\mathbf{u}_i^{(4)}\cdot\mathbf{n}^{(4)} = 0 \qquad , \qquad \mathbf{u}_i^{(4)}\cdot\mathbf{u}_j^{(4)} = 0 \qquad . \qquad \text{(C9)}$$

The normal $\mathbf{n}$ to the hypersurface $F(x, y, z, v/f) = 0$ reads,

$$\mathbf{n} = \frac{\nabla F}{\sqrt{\nabla F \cdot \nabla F}} = \frac{\left[\begin{array}{cccc} -v_x & -v_y & -v_z & f \end{array}\right]}{\sqrt{f^2 + \nabla v \cdot \nabla v}} \qquad , \qquad F = v - v(x, y, z) \qquad . \qquad \text{(C10)}$$

We demonstrate the orthogonality of equation C9 in Appendix F, where we provide an example of computing the principal curvatures and their directions for a point of a 3D curved hypersurface in 4D space.

## APPENDIX D. RAISING THE MATRIX TO A FRACTIONAL POWER

A symmetric matrix allows computing any scalar function (e.g., logarithm, sine, square root, etc.) of this matrix (e.g., Higham, 2008; Gantmacher, 2012). Since raising to a fractional power requires a positive argument, the matrix should be not only symmetric but also positive definite, which is the case for matrix $\mathbf{A}$ of the first fundamental form. To compute the function of a matrix, we first diagonalize the matrix (by performing a rotation), apply the scalar function to each diagonal component, and then rotate the result back to the original frame. The columns of the rotation matrix $\mathbf{V}$ are the eigenvectors of $\mathbf{A}$. The matrix of the first fundamental form of dimension $n$,

$$\mathbf{A} = \mathbf{I} + \frac{\nabla v \otimes \nabla v}{f^2} \qquad , \qquad (D1)$$

has an eigenvalue 1 of both algebraic and geometric multiplicity $n-1$, and a single simple eigenvalue,

$$\lambda = \frac{f^2 + \nabla v \cdot \nabla v}{f^2} \qquad . \qquad (D2)$$

The eigenvector corresponding to the single simple eigenvalue is the gradient $\nabla v$ of a scalar parameter whose curvatures are being studied. The two other eigenvectors (in the 3D scalar volume case) belong to the plane normal to the gradient; these two vectors are also normal to each other, but their orientation still contains a rotational degree of freedom – they can rotate about the gradient direction axis. The diagonal matrix (for a 3D hypersurface in 4D space) reads,

$$\mathbf{D} = \mathbf{V}^T \mathbf{A} \mathbf{V} = \begin{bmatrix} 1 & 1 & \dfrac{f^2 + \nabla v \cdot \nabla v}{f^2} \end{bmatrix} \qquad . \tag{D3}$$

After raising to power $-1/2$, we obtain

$$\mathbf{D}^{-1/2} = \left( \mathbf{V}^T \mathbf{A} \mathbf{V} \right)^{-1/2} = \begin{bmatrix} 1 & 1 & \dfrac{f}{\sqrt{f^2 + \nabla v \cdot \nabla v}} \end{bmatrix} \qquad . \tag{D4}$$

Eventually, we perform the backward rotation to the original frame,

$$\mathbf{A}^{-1/2} = \mathbf{V} \mathbf{D}^{-1/2} \mathbf{V}^T = \mathbf{V} \left( \mathbf{V}^T \mathbf{A} \mathbf{V} \right)^{-1/2} \mathbf{V}^T \qquad . \tag{D5}$$

This operation can also be performed analytically, resulting in,

$$\mathbf{A}^{-1/2} = \left( \mathbf{I} + \frac{\nabla v \otimes \nabla v}{f^2} \right)^{-1/2} = \mathbf{I} - \frac{\nabla v \otimes \nabla v}{\nabla v \cdot \nabla v} \cdot \frac{\sqrt{f^2 + \nabla v \cdot \nabla v} - f}{\sqrt{f^2 + \nabla v \cdot \nabla v}} \qquad . \tag{D6}$$

In the case of a vanishing gradient, $\mathbf{A} = \mathbf{I}$ and $\mathbf{A}^{-1/2} = \mathbf{I}$.

Rationalization

Except for very sharp transition zones, the scaling factor may essentially exceed the magnitude of the parameter gradient, $f \gg \sqrt{\nabla v \cdot \nabla v}$, and the square roots in equation D6 can be rationalized,

$$\frac{\nabla v \otimes \nabla v}{\nabla v \cdot \nabla v} \cdot \frac{\sqrt{f^2 + \nabla v \cdot \nabla v} - f}{\sqrt{f^2 + \nabla v \cdot \nabla v}} = \frac{\nabla v \otimes \nabla v}{2 f^2} + O \left( \nabla v^4 \right) \qquad , \tag{D7}$$

so that,

$$\mathbf{A}^{-1/2} \approx \mathbf{I} - \frac{\nabla v \otimes \nabla v}{2f^2} \qquad . \qquad (D8)$$

Note that this approximation is only valid for the condition above (slow spatial change of the parameter), and we recommend computing the exact matrix $\mathbf{A}^{-1/2}$ with equation D6; approximation D8 should only be considered in order to better understand the physical meaning of $\mathbf{A}^{-1/2}$.

## APPENDIX E. CHARACTERISTIC EQUATIONS FOR PRINCIPAL CURVATURES

The coefficients of the characteristic equations represent the effective curvatures, while their roots are the effective curvatures.

### 2D Surface in 3D Space

The characteristic quadratic equation reads,

$$k^2 - Ak + B = 0 \qquad , \qquad (E1)$$

where $A$ is the mean curvature doubled,

$$A = k_1 + k_2 = -\frac{\nabla v \cdot \nabla\nabla v \cdot \nabla v - \left(f^2 + \nabla v \cdot \nabla v\right)\mathrm{tr}\nabla\nabla v}{\left(f^2 + \nabla v \cdot \nabla v\right)^{3/2}} \qquad , \qquad (E2)$$

and $B$ is the Gaussian curvature,

$$B = k_1 k_2 = \frac{f^2 \det \nabla\nabla v}{\left(f^2 + \nabla v \cdot \nabla v\right)^2} \qquad . \qquad \text{(E3)}$$

## 3D Surface in 4D Space

The characteristic cubic equation reads,

$$k^3 - A k^2 + B k - C = 0 \qquad , \qquad \text{(E4)}$$

where, according to Vieta's theorem,

$$\underbrace{A = k_1 + k_2 + k_3}_{\substack{A/3 \text{ is the mean} \\ \text{curvature}}} \quad , \quad \underbrace{B = k_1 k_2 + k_2 k_3 + k_3 k_1}_{\substack{B/3 \text{ is the normalized} \\ \text{scalar curvature}}} \quad , \quad \underbrace{C = k_1 k_2 k_3}_{\substack{\text{Gaussian} \\ \text{curvature}}} \qquad . \qquad \text{(E5)}$$

Coefficient $A$ is the mean curvature tripled,

$$A = k_1 + k_2 + k_3 = -\frac{\nabla v \cdot \nabla\nabla v \cdot \nabla v - \left(f^2 + \nabla v \cdot \nabla v\right)\operatorname{tr}\nabla\nabla v}{\left(f^2 + \nabla v \cdot \nabla v\right)^{3/2}} \qquad , \qquad \text{(E6)}$$

coefficient $B$ is the scalar curvature of the 3D hypersurface in 4D Euclidean space (Suceava, 2014),

$$B = k_1 k_2 + k_2 k_3 + k_3 k_1 = \frac{\nabla v \cdot \nabla\nabla^* v \cdot \nabla v + f^2 \operatorname{tr}\nabla\nabla v^*}{\left(f^2 + \nabla v \cdot \nabla v\right)^2} \qquad , \qquad \text{(E7)}$$

and coefficient $C$ is the Gaussian curvature,

$$C = k_1 k_2 k_3 = \frac{f^2 \det \nabla\nabla v}{\left(f^2 + \nabla v \cdot \nabla v\right)^{5/2}} \qquad . \qquad (E8)$$

Matrix $\nabla\nabla v^*$ in equation E7 is the adjoint matrix of the scalar field Hessian consisting of the cofactors of the original matrix and then transposed. (In this particular case, the transpose operator can be omitted due to symmetry). The adjoint matrix may be used to compute the inverse matrix. Therefore, for an arbitrary square matrix $\mathbf{P}$, in the case of a non-vanishing determinant, the adjoint matrix can be arranged in a simpler form, $\mathbf{P}^* = \mathbf{P}^{-1} \det \mathbf{P}$. The coefficients of quadratic and cubic polynomials in equations D1 and D4, respectively, are the curvature invariants referred to as effective curvatures.

## APPENDIX F. COMPUTING PRINCIPAL CURVATURES AND THEIR DIRECTIONS

In this appendix, we demonstrate the computation of the principal curvatures and their corresponding principal directions for a single grid node of the 3D scalar field. Consider, for example, a velocity volume as the scalar field, in this case, the scaling factor $f$ is the dominant frequency. Assume the following values of the velocity gradient $\nabla v$ and Hessian $\nabla\nabla v$ components at the given point,

$$\nabla v = \begin{bmatrix} +0.54 \\ -0.37 \\ +0.38 \end{bmatrix} \frac{1}{\text{s}} \ , \quad \nabla\nabla v = \begin{bmatrix} +0.48 & -0.24 & +0.37 \\ -0.24 & +0.56 & -0.15 \\ +0.37 & -0.15 & +0.68 \end{bmatrix} \frac{1}{\text{km}\cdot\text{s}} \ , \quad f = 56 \ \text{Hz} \qquad . \qquad (F1)$$

Computing the matrices of the first and second fundamental forms of the hypersurface, and applying equations 2 and 3, respectively, we obtain,

$$\mathbf{A} = \begin{bmatrix} +1.000092985 & -6.371173469 \cdot 10^{-5} & +6.543367347 \cdot 10^{-5} \\ -6.371173469 \cdot 10^{-5} & +1.000043654 & -4.483418367 \cdot 10^{-5} \\ +6.543367347 \cdot 10^{-5} & -4.483418367 \cdot 10^{-5} & +1.000046046 \end{bmatrix} \qquad . \qquad \text{(F2)}$$

Components of matrix $\mathbf{A}$ are unitless; the units of matrix $\mathbf{B}$ components are reciprocals of distance,

$$\mathbf{B} = \begin{bmatrix} +8.570645743 \cdot 10^{-3} & -4.285322872 \cdot 10^{-3} & +6.606539427 \cdot 10^{-3} \\ -4.285322872 \cdot 10^{-3} & +9.999086700 \cdot 10^{-3} & -2.678326795 \cdot 10^{-3} \\ +6.606539427 \cdot 10^{-3} & -2.678326795 \cdot 10^{-3} & +1.214174814 \cdot 10^{-2} \end{bmatrix} \frac{1}{\text{km}} \qquad . \qquad \text{(F3)}$$

Next, we compute $\mathbf{A}^{-1/2}$ applying equation D6,

$$\mathbf{A}^{-1/2} = \begin{bmatrix} +9.999535140 \cdot 10^{-1} & +3.185150332 \cdot 10^{-5} & -3.271235476 \cdot 10^{-5} \\ +3.185150332 \cdot 10^{-5} & +9.999781758 \cdot 10^{-1} & +2.241402086 \cdot 10^{-5} \\ -3.271235476 \cdot 10^{-5} & +2.241402086 \cdot 10^{-5} & +9.999769802 \cdot 10^{-1} \end{bmatrix} \qquad . \qquad \text{(F4)}$$

We then compute matrix $\mathbf{C}$, applying equation 6,

$$\mathbf{C} = \mathbf{A}^{-1/2}\,\mathbf{B}\,\mathbf{A}^{-1/2} =$$
$$\begin{bmatrix} +8.569143775 \cdot 10^{-3} & -4.284203028 \cdot 10^{-3} & +6.605221379 \cdot 10^{-3} \\ -4.284203028 \cdot 10^{-3} & +9.998257242 \cdot 10^{-3} & -2.677359850 \cdot 10^{-3} \\ +6.605221379 \cdot 10^{-3} & -2.677359850 \cdot 10^{-3} & +1.214063688 \cdot 10^{-2} \end{bmatrix} \frac{1}{\text{km}} \qquad . \qquad \text{(F5)}$$

The eigenvalues of matrix $\mathbf{C}$ are the minimum, medium-valued, and maximum curvatures. In this particular case, they are all positive (this is not a must, of course).

$$k = \begin{bmatrix} +2.939084202 \cdot 10^{-3} & +8.188073253 \cdot 10^{-3} & +1.958088044 \cdot 10^{-2} \end{bmatrix} \frac{1}{\text{km}} \qquad . \qquad \text{(F6)}$$

Applying equations E6, E7, and E8, we can also compute the coefficients of the characteristic cubic equation for the principal curvatures,

$$A = 3.0708038 \times 10^{-2} \text{ km}^{-1}, \quad B = 2.4194498 \times 10^{-4} \text{ km}^{-2}, \quad C = 4.7122244 \times 10^{-7} \text{ km}^{-3}. \quad (F7)$$

Recall that the coefficients in equation F7 represent the effective curvatures. The solution of the cubic equation E4 coincides, of course, with the results in equation F6.

The eigenvectors $\mathbf{w}_i$ of matrix $\mathbf{C}$ represent the $\mathbf{z}$-space principal directions,

$$\begin{bmatrix} \mathbf{w}_1 & \mathbf{w}_2 & \mathbf{w}_3 \\ +0.8133500630 & +0.01496272677 & -0.5815821454 \\ +0.3059437213 & +0.8392766106 & +0.4494587972 \\ -0.4948334209 & +0.5434987468 & -0.6780479317 \end{bmatrix} \quad \text{eigenvectors in space} \quad \mathbf{z}. \quad (F8)$$

These eigenvectors are orthogonal to each other, $\mathbf{w}_i \cdot \mathbf{w}_j = 0, \ i \neq j$. We convert these eigenvectors from $\mathbf{z}$-space to our original $\mathbf{r}$-space, applying equation C8. This transform ruins the normalization, and we then re-normalize the eigenvectors to the unit length. This leads to the following principal directions,

$$\begin{bmatrix} \mathbf{u}_1 & \mathbf{u}_2 & \mathbf{u}_3 \\ +0.8133406537 & +0.01497100627 & -0.5815691101 \\ +0.3059527879 & +0.8392721836 & +0.4494542914 \\ -0.4948432808 & +0.5435053549 & -0.6780620990 \end{bmatrix} \quad \text{eigenvectors in space} \quad \mathbf{r}. \quad (F9)$$

The eigenvectors in space $\mathbf{r}$ are not orthogonal to each other, $\mathbf{u}_i \cdot \mathbf{u}_j \neq 0, \ i \neq j$. Should they be? Of course – this is the nature of the principal directions for symmetric matrices. However, they are orthogonal in the full 4D space, while in equation F9 only $x, y, z$ components of the eigenvectors are presented, and the altitude $v$-component is missing. Next, we compute the fourth component (in this case, the velocity field component) for each of the three eigenvectors and demonstrate the orthogonality.

The principal directions belong to the tangent plane of the curved surface (in 3D space) or in our case, the curved hypersurface (in 4D space). Any direction in the tangent hyperplane is orthogonal to the surface normal, $\mathbf{n}$. This normal, in turn, is the gradient of the 4D *vanishing* surface function $F(x, y, z, \lambda)$, where $\lambda = v / f$ is the scaled scalar field (in our example – the wavelength),

$$v = v(x, y, z) \qquad \rightarrow \qquad F = \lambda f - v(x, y, z) \qquad . \qquad \text{(F10)}$$

The gradient of the scalar function $F$ reads,

$$\nabla F = \begin{bmatrix} \dfrac{\partial F}{\partial x} & \dfrac{\partial F}{\partial y} & \dfrac{\partial F}{\partial z} & \dfrac{\partial F}{\partial \lambda} \end{bmatrix} = \begin{bmatrix} -v_x & -v_y & -v_z & +f \end{bmatrix} \qquad . \qquad \text{(F11)}$$

Normalizing this vector to the unit length, we obtain,

$$\mathbf{n} = \frac{\nabla F}{|\nabla F|} = \frac{\begin{bmatrix} -v_x & -v_y & -v_z & +f \end{bmatrix}}{\sqrt{f^2 + \nabla v \cdot \nabla v}} \qquad . \qquad \text{(F12)}$$

For the data in this example,

$$\mathbf{n} = \begin{bmatrix} -9.6419765 \cdot 10^{-3} & +6.6065394 \cdot 10^{-3} & -6.7850945 \cdot 10^{-3} & +9.9990867 \cdot 10^{-1} \end{bmatrix} \quad . \quad \text{(F13)}$$

The last component of the normal has proven to be much larger than the first three because the dominant frequency essentially exceeds the velocity gradient length for the data used in this example.

As mentioned, any vector in the tangent plane is orthogonal to the surface normal,

$$\mathbf{u}_i^{(4)} \cdot \mathbf{n} = 0 \quad \rightarrow \quad u_{i,x}n_x + u_{i,y}n_y + u_{i,z}n_z + u_{i,v}n_v = 0 \qquad , \qquad \text{(F14)}$$

where the superscript number 4 emphasizes the dimension of the hyperspace. Similarly, for a 2D surface in 3D space, the azimuths of the principal directions for the minimum and maximum principal curvatures differ by $90^o$ in the tangent plane. This is not so for the projections of these azimuthal directions onto the horizontal plane, however. By analogy with a 2D surface in 3D space, one can say that $\mathbf{u}^{(4)}$ is the full principal direction for a 3D surface in 4D space, while $\mathbf{u}^{(3)}$ is only the azimuth of the principal direction, with the "vertical" (altitude) component missing. In four-dimensional geometry (e.g., Forsyth, 1930; Manning, 2007), a hyperplane (which is often simply called a plane) is defined by three intersecting lines. In particular, the Cartesian sub-space $xyz$ (for a fixed $v$) can be considered the "horizontal" hyperplane, and the value axis $v$ can be assumed "vertical".

Equation F14 makes it possible to establish the missing component,

$$u_{i,v} = -\frac{u_{i,x}n_x + u_{i,y}n_y + u_{i,z}n_z}{n_v} = +\frac{u_{i,x}v_x + u_{i,y}v_y + u_{i,z}v_z}{f} = \frac{\mathbf{u}_i^{(3)} \cdot \nabla v}{f} \qquad . \qquad \text{(F15)}$$

Adding the fourth component ruins the normalization, so we re-normalize the eigenvectors $\mathbf{u}_i^{(4)}$ to the unit length. Eventually, we obtain the final results,

|  | Minimum curvature | Medium curvature | Maximum curvature |
|---|---|---|---|

$$k_{\min} = +2.9390842 \cdot 10^{-3}\ \text{km}^{-1} \qquad k_{\text{med}} = +8.1880733 \cdot 10^{-3}\ \text{km}^{-1} \qquad k_{\max} = +1.9580880 \cdot 10^{-2}\ \text{km}^{-1}$$

$$u_{\min,x} = +8.1333819 \cdot 10^{-1} \qquad u_{\text{med},x} = +1.4970984 \cdot 10^{-2} \qquad u_{\max,x} = +5.8151861 \cdot 10^{-1}$$

$$u_{\min,y} = +3.0595186 \cdot 10^{-1} \qquad u_{\text{med},y} = +8.3927095 \cdot 10^{-1} \qquad u_{\max,y} = -4.4941527 \cdot 10^{-1}$$

$$u_{\min,z} = -4.9484178 \cdot 10^{-1} \qquad u_{\text{med},z} = +5.4350456 \cdot 10^{-1} \qquad u_{\max,z} = +6.7800322 \cdot 10^{-1}$$

$$u_{\min,v} = +2.4635814 \cdot 10^{-3} \qquad u_{\text{med},v} = -1.7127534 \cdot 10^{-3} \qquad u_{\max,v} = +1.3177588 \cdot 10^{-2}$$

$$\text{(F16)}$$

and make sure that $\mathbf{u}_i^{(4)} \cdot \mathbf{u}_j^{(4)} = 0$, $i \neq j$. Since the three eigenvectors and the normal to the curved hypersurface are all mutually orthogonal, they constitute a local Cartesian frame of reference. The square matrix of dimension 4,

$$\mathbf{R} = \begin{bmatrix} \mathbf{u}_{\min} & \mathbf{u}_{\text{med}} & \mathbf{u}_{\max} & \mathbf{n} \end{bmatrix}, \qquad \text{(F17)}$$

where each entry in brackets is a column, represents the local-to-global rotation matrix in 4D space. We make sure that $\det \mathbf{R} = +1$ in order to keep the "right frame" after rotation; otherwise, in the case of $\det \mathbf{R} = -1$, we negate the signs for all components of one eigenvector,

$$\mathbf{R} = \begin{bmatrix} +8.1333819 \cdot 10^{-1} & +1.4970984 \cdot 10^{-2} & +5.8151861 \cdot 10^{-1} & -9.6419765 \cdot 10^{-3} \\ +3.0595186 \cdot 10^{-1} & +8.3927095 \cdot 10^{-1} & -4.4941527 \cdot 10^{-1} & +6.6065394 \cdot 10^{-3} \\ -4.9484178 \cdot 10^{-1} & +5.4350456 \cdot 10^{-1} & +6.7800322 \cdot 10^{-1} & -6.7850945 \cdot 10^{-3} \\ +2.4635814 \cdot 10^{-3} & -1.7127534 \cdot 10^{-3} & +1.3177588 \cdot 10^{-2} & +9.9990867 \cdot 10^{-1} \end{bmatrix}. \quad \text{(F18)}$$

The eigenvalues of the 4D rotation matrix are normally two pairs of complex-conjugate numbers with magnitude 1. (The eigenvalues of the 3D rotation matrix represent a pair of complex-conjugate numbers with the magnitude 1 and the real number 1). As in a 3D case, the 4D rotation matrix is unitary: its transpose is also its inverse, where for an arbitrary 4D vector $\mathbf{v}$, the local-to-global and global-to-local transformations read,

$$\mathbf{v}_{\text{glb}} = \mathbf{R}\,\mathbf{v}_{\text{loc}} \quad , \qquad \mathbf{v}_{\text{loc}} = \mathbf{R}^T \mathbf{v}_{\text{glb}} \qquad . \tag{F19}$$

## APPENDIX G. SPHERICAL ANGLES FOR GEOMETRY OF FOUR DIMENSIONS

Since the principal directions are normalized to the unit length, it is suitable to represent them in the form of the polar angles in the spherical frame of reference, where the radius $R = 1$. The spherical angles of the four-dimensional geometry are discussed below.

In the case of a regular 3D geometry, the spherical angles are the dip, $0 \le \theta \le \pi$, and the azimuth, $0 \le \varphi < 2\pi$, where the transform between the Cartesian and spherical coordinates reads,

$$
\begin{aligned}
x &= R\sin\theta\sin\varphi \\
y &= R\sin\theta\cos\varphi \\
z &= R\cos\theta
\end{aligned}
\qquad
\begin{aligned}
\theta &= \arctan\left(\sqrt{x^2 + y^2}, z\right) = \arccos\frac{z}{R} \\
\varphi &= \arctan\left(y, x\right) \\
R &= \sqrt{x^2 + y^2 + z^2}
\end{aligned}
\qquad . \tag{G1}
$$

A similar representation exists in 4D, with the zenith angle, $0 \le \psi \le \pi$, the dip angle, $0 \le \theta \le \pi$, and the azimuth angle $0 \le \varphi < 2\pi$, where the transform reads,

$$
\begin{aligned}
x &= R\sin\psi\sin\theta\cos\varphi \\
y &= R\sin\psi\sin\theta\sin\varphi \\
z &= R\sin\psi\cos\theta \\
\lambda &= R\cos\psi
\end{aligned}
\qquad
\begin{aligned}
\psi &= \arctan\left(\sqrt{x^2 + y^2 + z^2}, \lambda\right) = \arccos\frac{\lambda}{R} \\
\theta &= \arctan\left(\sqrt{x^2 + y^2}, z\right) = \arccos\frac{z}{\sqrt{x^2 + y^2 + z^2}} \\
\varphi &= \arctan\left(y, x\right) \\
R &= \sqrt{x^2 + y^2 + z^2 + \lambda^2}
\end{aligned}
\qquad . \tag{G2}
$$

In the case $\lambda = v/f \gg \sqrt{\nabla v \cdot \nabla v}$ (i.e., apart from the transition zones), the additional 4D zenith angle $\psi$ is close to $\pi/2$ since the tangent hyperplane is almost normal to the scalar field $v$ (e.g., velocity) coordinate axis. Note that for a fixed zenith angle $\psi$, the 3D sphere of radius $R$ in 4D space reduces to a 2D sphere of radius $R\sin\psi$ in regular 3D space.

It may be suitable to store the four-dimensional vectors (either in the spherical or Cartesian frame) in the form of quaternions,

$$q = R\exp\left(\tau\psi\right) = R\cos\psi + R\boldsymbol{\tau}\sin\psi \qquad , \qquad (G3)$$

where $\boldsymbol{\tau}$ is the unit imaginary quaternion,

$$\boldsymbol{\tau} = \sin\theta\cos\varphi\,\mathbf{i} + \sin\theta\sin\varphi\,\mathbf{j} + \cos\theta\,\mathbf{k} \;, \qquad \tau^2 = -1 \qquad . \qquad (G4)$$

In the 4D Cartesian frame,

$$\begin{aligned} q = {} & x\,\mathbf{i} + y\,\mathbf{j} + z\,\mathbf{k} + v/f = \\ & R\left(\sin\psi\sin\theta\cos\varphi\,\mathbf{i} + \sin\psi\sin\theta\sin\varphi\,\mathbf{j} + \sin\psi\cos\theta\,\mathbf{k} + \cos\psi\right) \quad . \end{aligned} \qquad (G5)$$

For $R = 1$, the quaternion $q$ becomes a versor (quaternion of the unit length).

## APPENDIX H. VOLUME PRE-CONDITIONING AND COMPUTING DERIVATIVES

Consider an unsmooth scalar field (in this appendix we refer to the velocity field volume $v_{\mathrm{o}} = v_{\mathrm{o}}\left(x, y, z, t = 0\right)$, where $t$ belongs to the scale of the smoothing operator, rather than represents the actual time). We first smooth the scalar field and then compute its gradient and Hessian components in the Fourier space. For this, we perform a 3D FFT (Fast Fourier Transform) of the space-domain scalar field volume to the wave-number domain: Real-to-complex in the vertical direction and complex-to-complex in the two lateral directions. Accordingly, the data is stored vs. non-negative vertical wavenumbers and lateral wavenumbers of any sign. The smoothing is modeled with the heat transfer equation for a medium with the temperature distribution $v(x, y, z, t)$ with the homogeneous and isotropic thermal conductivity $\alpha$,

$$\frac{\partial v(x, y, z, t)}{\partial t} = \alpha \left( v_{xx} + v_{yy} + v_{zz} \right) \qquad . \qquad (H1)$$

In the Fourier domain, where $\bar{v}$ is the image of the scalar field, this partial differential equation becomes an ordinary differential equation,

$$\frac{d\bar{v}}{dt} = -\alpha \left( k_x^2 + k_y^2 + k_z^2 \right) \bar{v} \qquad , \qquad (H2)$$

with the analytic solution,

$$\bar{v} = \bar{v}_o \exp \left[ -\alpha t \left( k_x^2 + k_y^2 + k_z^2 \right) \right] \qquad , \qquad (H3)$$

where $\bar{v}_o$ is the Fourier image of the initial (unsmooth) scalar field. Parameter $\alpha t \equiv \tau$ is called the scale and it characterizes the degree of smoothing; it has the units of area. Zero value of $\tau$ corresponds to an unsmooth scalar field. After the volume is smoothed, we compute the derivatives in the Fourier domain,

$$\nabla \bar{v} = i \mathbf{k} \, \bar{v} \quad , \qquad \nabla \nabla \bar{v} = -\mathbf{k} \otimes \mathbf{k} \, \bar{v} \quad , \qquad \text{where} \qquad \mathbf{k} = \begin{bmatrix} k_x & k_y & k_z \end{bmatrix} \qquad . \qquad (H4)$$

With these notations, the 3D smoothing equation H3 reduces to,

$$\overline{v}(\mathbf{k}, \tau) = \overline{v}_o(\mathbf{k}) \exp(-\mathbf{k} \cdot \mathbf{k} \ \tau) \qquad . \qquad \text{(H5)}$$

As one can see from this equation, the magnitudes of the scalar field which are related to the high wavenumbers are damped out exponentially, and the volume becomes smooth. Recall that,

$$k_x = \frac{2\pi i_x}{\Delta x \, n_x^{\text{FFT}}}, \quad -\frac{n_x^{\text{FFT}}}{2} < i_x \leq +\frac{n_x^{\text{FFT}}}{2} \qquad , \qquad \text{(H6)}$$

and similar formulae exist in $y$ and $z$ dimensions. As mentioned, the negative vertical wavenumbers are not considered since the Fourier transform in the vertical direction is real-to-complex, $\overline{v}(i_x, i_y, -i_z) = \overline{v}^*(i_x, i_y, i_z)$, where the asterisk means complex conjugate.

Considering equation H6, $k_x, k_y, k_z$ are the wavenumbers, $\Delta x, \Delta y, \Delta z$ are the scalar field (e.g., velocity) grid resolutions, $n_x^{\text{FFT}}, n_y^{\text{FFT}}, n_z^{\text{FFT}}$, are the numbers of samples in the corresponding dimensions (including the pad samples added to provide periodicity of the data and achieve the supported Fourier numbers), and $i_x, i_y, i_z$ are the wavenumber indices, which may be zero, positive or negative for $x$ and $y$, and only zero or positive for $z$. The Nyquist wavenumber index $i = n^{\text{FFT}} / 2$ exists only for an even number of samples in the corresponding dimension. Note that in all standard FFT routines, the data wavenumbers are arranged in the following sequence: DC, positive AC (starting from the smallest), Nyquist (if any), which may be considered both positive and negative, and negative AC (starting from the wavenumber with the largest absolute value). We apply a single direct Fourier transform for the scalar field, while the inverse transform is carried out for the scalar field and for all its gradient components. The latter are then refined/stabilized using the structure tensor technique (equation B4), and the Hessian components are computed in the Fourier space from the gradient components. The numerical operations are fully parallelized, including the forward and inverse FFT operations.

When a large-size scalar field volume needs to be processed, which (along with its derivatives and curvatures) does not fit the memory of the machine, the volume is split into a number of overlapping blocks (sub-volumes) (domain decomposition) where each sub-volume is processed independently, applying a tapered normalized weighted average in the overlapping regions.

## APPENDIX I. COMPUTATIONAL COMPLEXITY

In this appendix, we compare the computational complexities of computing the principal dip-based and hypersurface curvatures. Let $m$ be the required accuracy, i.e. the number of binary digits in a real number, where $m = 23$ and $m = 52$ for single and double precisions, respectively. A further estimate is performed for the double precision. Let $M$ be the complexity for a multiplication of real numbers with $m$ binary digits. Then, as demonstrated by Brent (1975), the complexity of an addition/subtraction can be ignored, the complexity of a division is $4M$, and that of a square root is around $6M$; the complexity of trigonometric functions (sine, cosine, and both of them) is $34M \log_2 m \approx 194M$, and complexity of their inverse is $17M \log_2^2 m \approx 553\, M$.

As explained in Appendix B, to compute the hypersurface curvatures, all the gradient and Hessian components of a scalar field are needed, while for the dip-based surface curvature, the second vertical derivative $\partial^2 v / \partial z^2$ can be skipped; however, we ignore its minor contribution into complexity. Computing coefficients $A, B$ for the quadratic equation E1 and coefficients $A, B, C$ for the cubic equation E4 requires around $28M$ and $65M$ floating point operations, respectively (equivalent multiplications). Solving a quadratic equation requires 4 multiplications, 2 divisions, and a square root, which leads to the complexity of $18M$. Solving a cubic equation with all real roots requires 18 multiplications, 4 divisions, a square root, 3 cosines, and an inverse cosine (Press et al., 2002), leading to an equivalent of $1175M$. Thus, after the first and second derivatives of a scalar field are established, we need $46M$ and $1240M$ operations to compute the principal curvatures of a surface and a hypersurface, respectively. As we see, the exact solution of a cubic eigenvalue equation includes expensive computations of trigonometric functions and their inverse.

Jacobi numerical eigenvalue algorithm (applied for symmetric matrices) has complexity $O(n^3)$ per iteration, where $n = 2, 3$ is the dimension of a matrix, for a surface and a hypersurface, respectively. The process normally converges after a small number of iterations; typically, the total labor is $12n^3M$ to $20n^3M$ operations (Press et al., 2002), which results in up to $20 \cdot 2^3 M = 160M$ for a surface and up to $20 \cdot 3^3 M = 540M$ for a hypersurface (per node of the volume grid). Thus, the numerical approach is more effective for computing the principal curvatures of a hypersurface.

## APPENDIX J. MECHANICAL AND ELECTRICAL ANALOGS

There is a direct analogy between the principal curvatures/directions of surfaces and hypersurfaces (equation 1) and the natural frequencies/modes of oscillations in mechanical and electrical systems. Indeed, for a system with multiple masses and springs or that with inductors and capacitors, with negligible damping or active electric resistance, the equations of oscillations are,

$$\mathbf{M}\frac{d^2\mathbf{x}}{dt^2}+\mathbf{Kx}=0 \qquad\qquad \mathbf{L}\frac{d^2\mathbf{i}}{dt^2}+\mathbf{C}^{-1}\mathbf{i}=0$$

$$\mathbf{x}=\mathbf{v}\cos\left(\omega_{\mathrm{o}}t+\psi\right) \qquad\qquad \mathbf{i}=\mathbf{v}\cos\left(\omega_{\mathrm{o}}t+\psi\right)$$

$$\omega_0^2\mathbf{Mv}\ =\ \mathbf{Kv} \qquad\qquad\quad \omega_0^2\mathbf{Lv}\ =\ \mathbf{C}^{-1}\mathbf{v}$$

$$\omega_0^2\mathbf{vMv}=\mathbf{vKv} \qquad\qquad \omega_0^2\mathbf{vLv}=\mathbf{vC}^{-1}\mathbf{v}$$

$$\omega_{\mathrm{o}}^2=\frac{\mathbf{vKv}}{\mathbf{vMv}} \qquad\qquad \omega_{\mathrm{o}}^2=\frac{\mathbf{vC}^{-1}\mathbf{v}}{\mathbf{vLv}} \qquad\qquad k=\frac{\mathbf{rBr}}{\mathbf{rAr}}$$

$$\omega_{\mathrm{o}}^2=\mathbf{wM}^{-1/2}\mathbf{KM}^{-1/2}\mathbf{w} \qquad \omega_{\mathrm{o}}^2=\mathbf{wL}^{-1/2}\mathbf{C}^{-1}\mathbf{L}^{-1/2}\mathbf{w} \qquad k=\mathbf{wA}^{-1/2}\mathbf{B}^{-1}\mathbf{A}^{-1/2}\mathbf{w}$$

$$,\quad (\mathrm{J}1)$$

where the first column of the equation set is related to mechanical systems, the second – to electrical, and the third one – to the surface or hypersurface curvatures. Systems with two degrees of freedom (DoF) are analogs of surfaces, and those with three or more DoF – hypersurfaces. The first line is the differential equation of motion, where in the left column, $\mathbf{M}$ is the mass matrix, $\mathbf{K}$ is the stiffness matrix, and $\mathbf{x}$ is the DoF displacement vector, while in the central column $\mathbf{L}$ is the inductance matrix (with the self-inductances on the diagonal and mutual inductances off the diagonal), $\mathbf{C}$ is the capacitance matrix (with self-capacitances on the diagonal, and either wanted or parasitic mutual capacities off the diagonal), and $\mathbf{i}$ is the DoF electric current vector. All matrices are symmetric and positive-definite because they are related to the energy of a linear system. Capacitance is the analog of compliance (rather than stiffness), and therefore matrix $\mathbf{C}$ appears inverted. In the second line of the equation set, we assume that a harmonic oscillation takes place, with the natural mode $\mathbf{v}$ (amplitude values of the DoF), the natural angular frequency $\omega_{\mathrm{o}}=2\pi f_{\mathrm{o}}$, and the initial phase $\psi$. (Indeed, the system may oscillate with any of the natural modes and with a superposition of all of them.) This leads to the third line of the equation set. Next, we multiply the result (from the left) by vector $\mathbf{v}$ to obtain the scalar value of the natural frequency squared, $\omega_0^2$ as a generalized Rayleigh quotient, and finally, it is converted to the standard quotient, where vector $\mathbf{w}$ is directly related to the mode vector $\mathbf{v}$. The mass and inductance matrices, $\mathbf{M}$ and $\mathbf{L}$, are analogs of the first fundamental tensor $\mathbf{A}$, while the stiffness and inverse capacitance matrices, $\mathbf{K}$ and $\mathbf{C}^{-1}$, are analogs of the second fundamental tensor $\mathbf{B}$. The natural frequencies squared are the stationary values of the quotient. Unlike the matrix for the principal curvatures (these curvatures may have any sign), the matrix for the natural frequencies is positive definite.

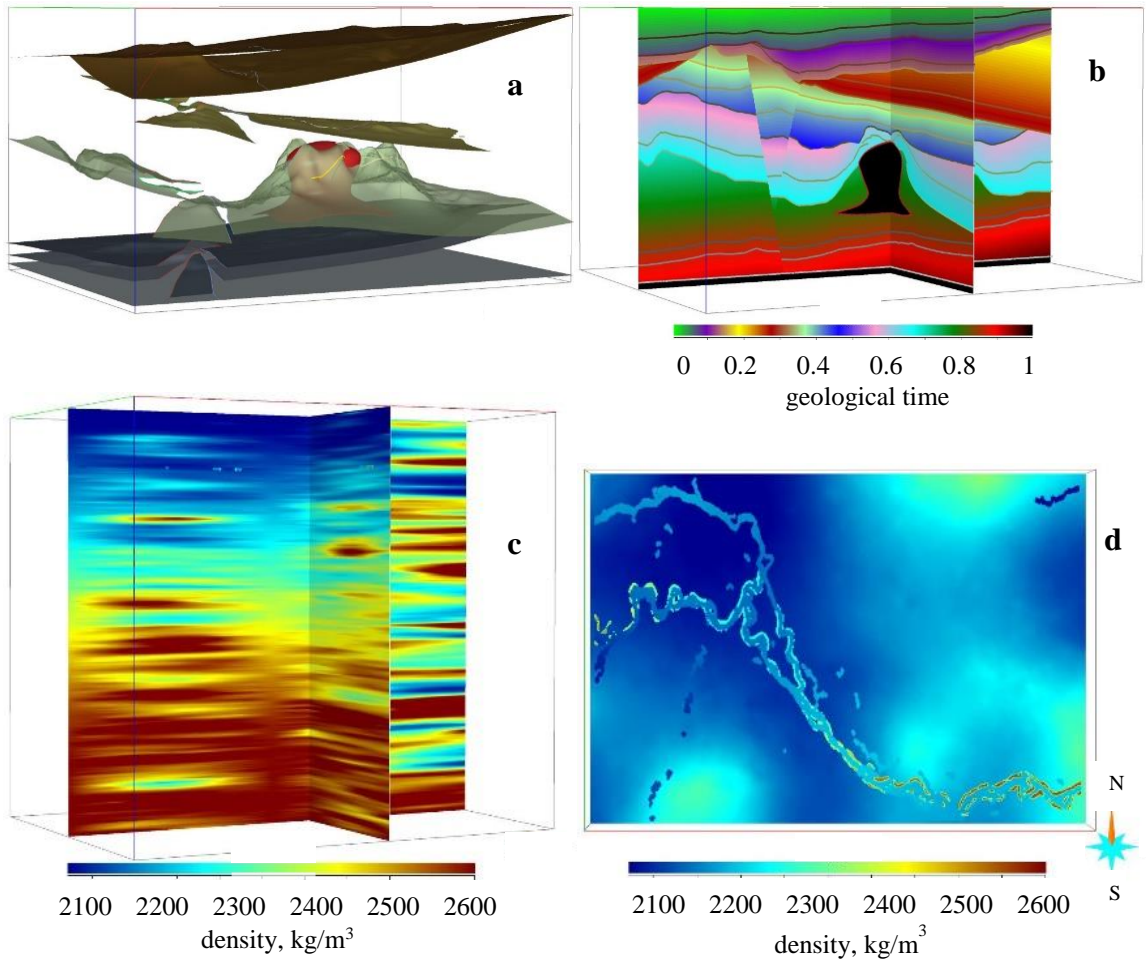

Figure 1. Modeling structure in the geological space and properties in UVT-transformed space

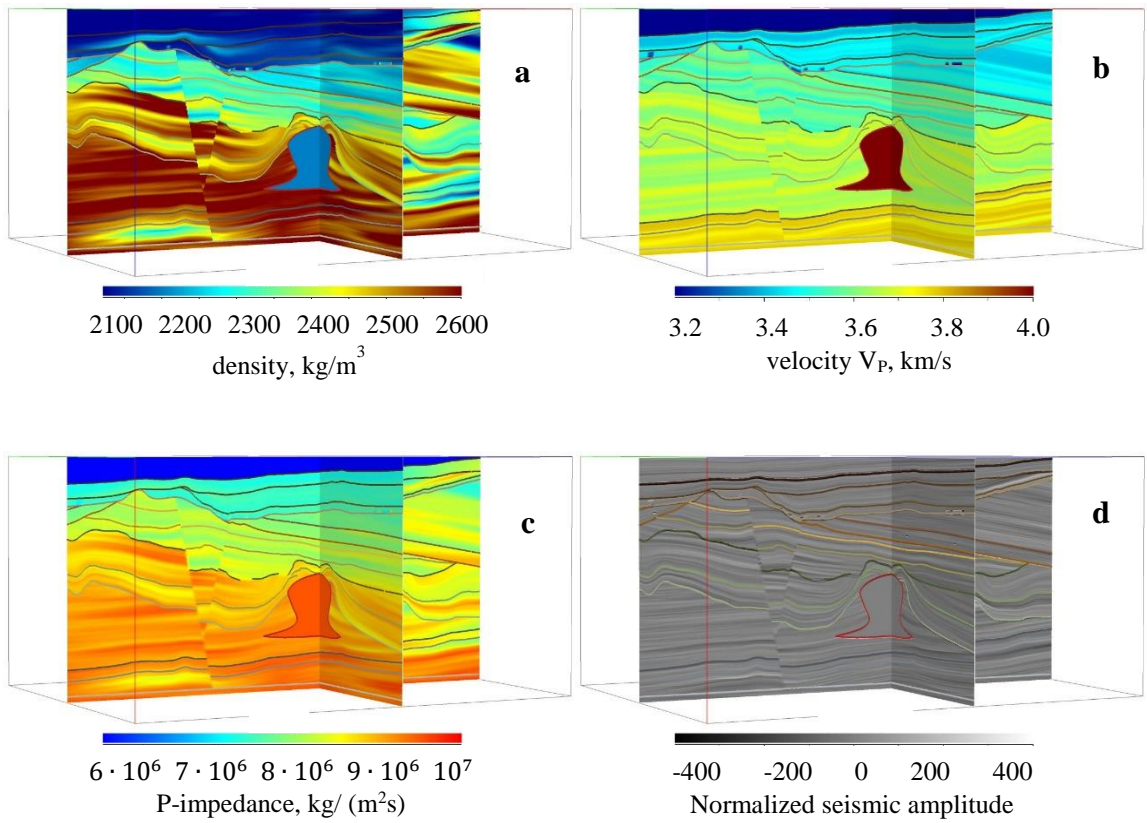

Figure 2.  Combining structure and geological features in geological space

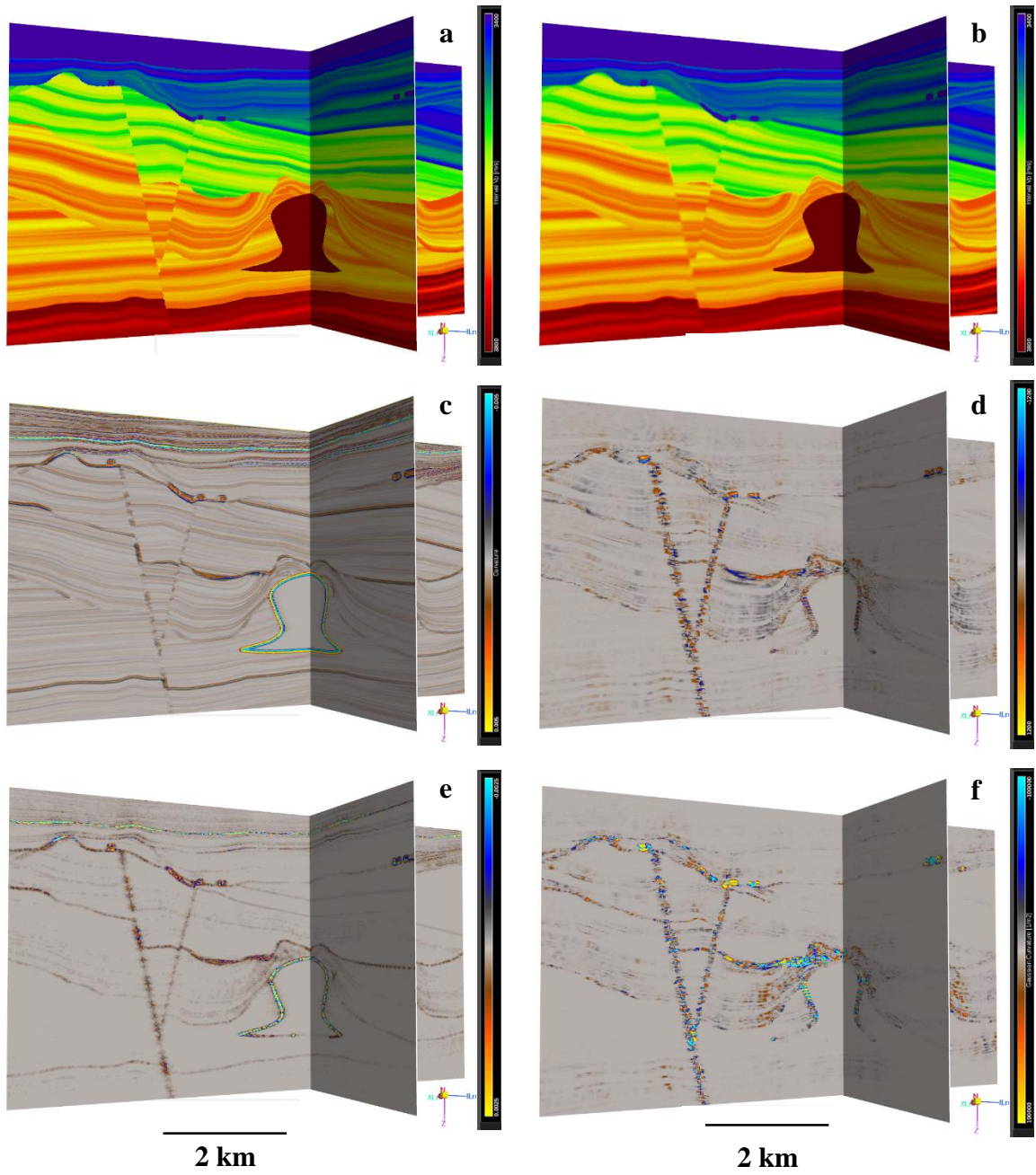

Figure 3. Volumetric curvatures of velocity field, 3D views, synthetic model

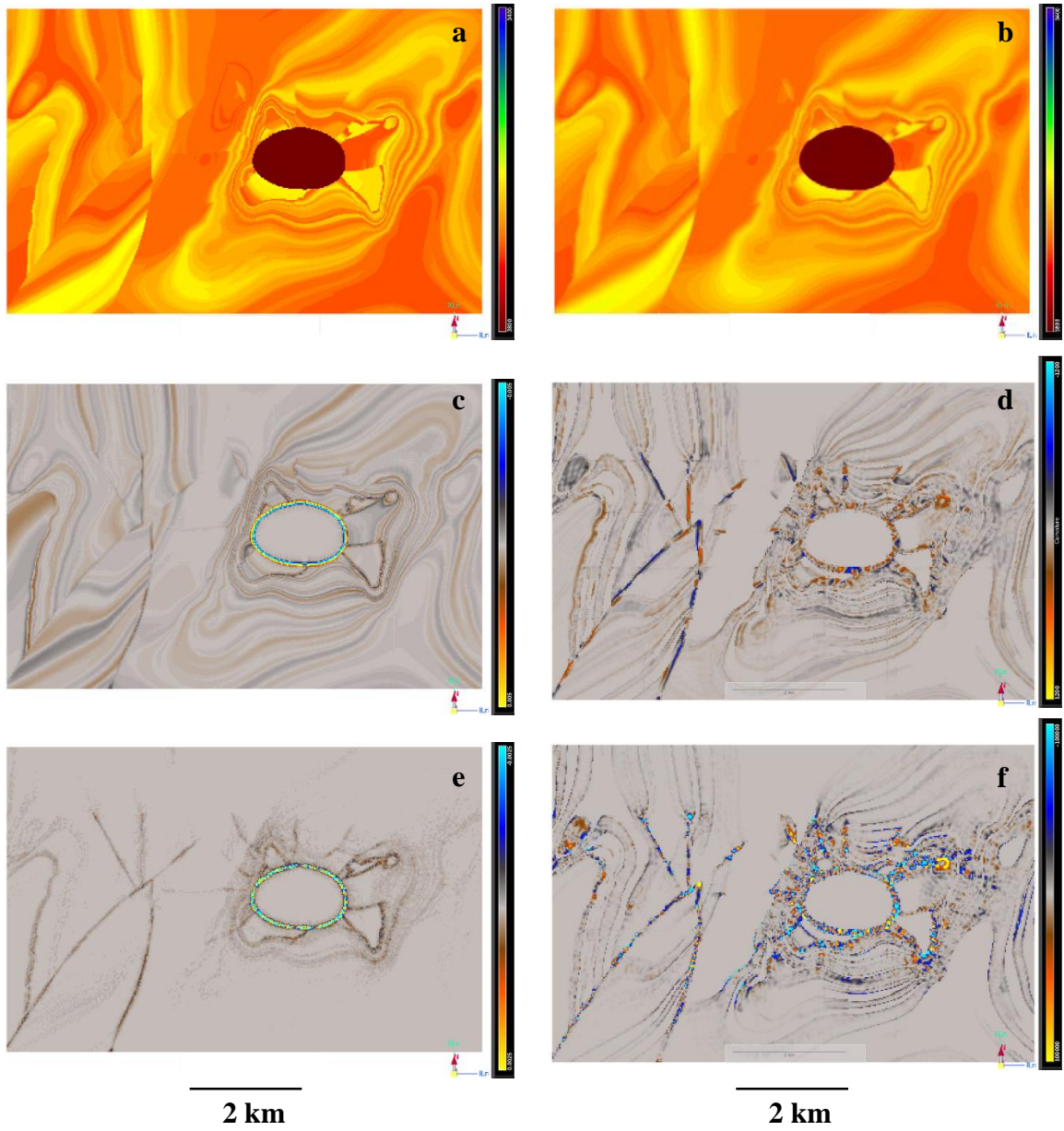

Figure 4. Volumetric curvatures of velocity field, horizontal slice views, synthetic model

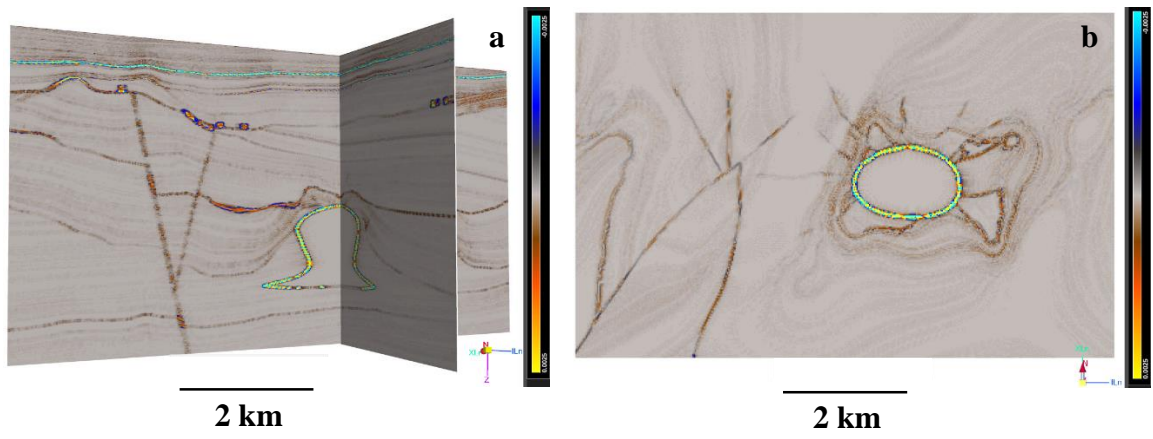

Figure 5. Scalar hypersurface curvature of velocity field, synthetic model

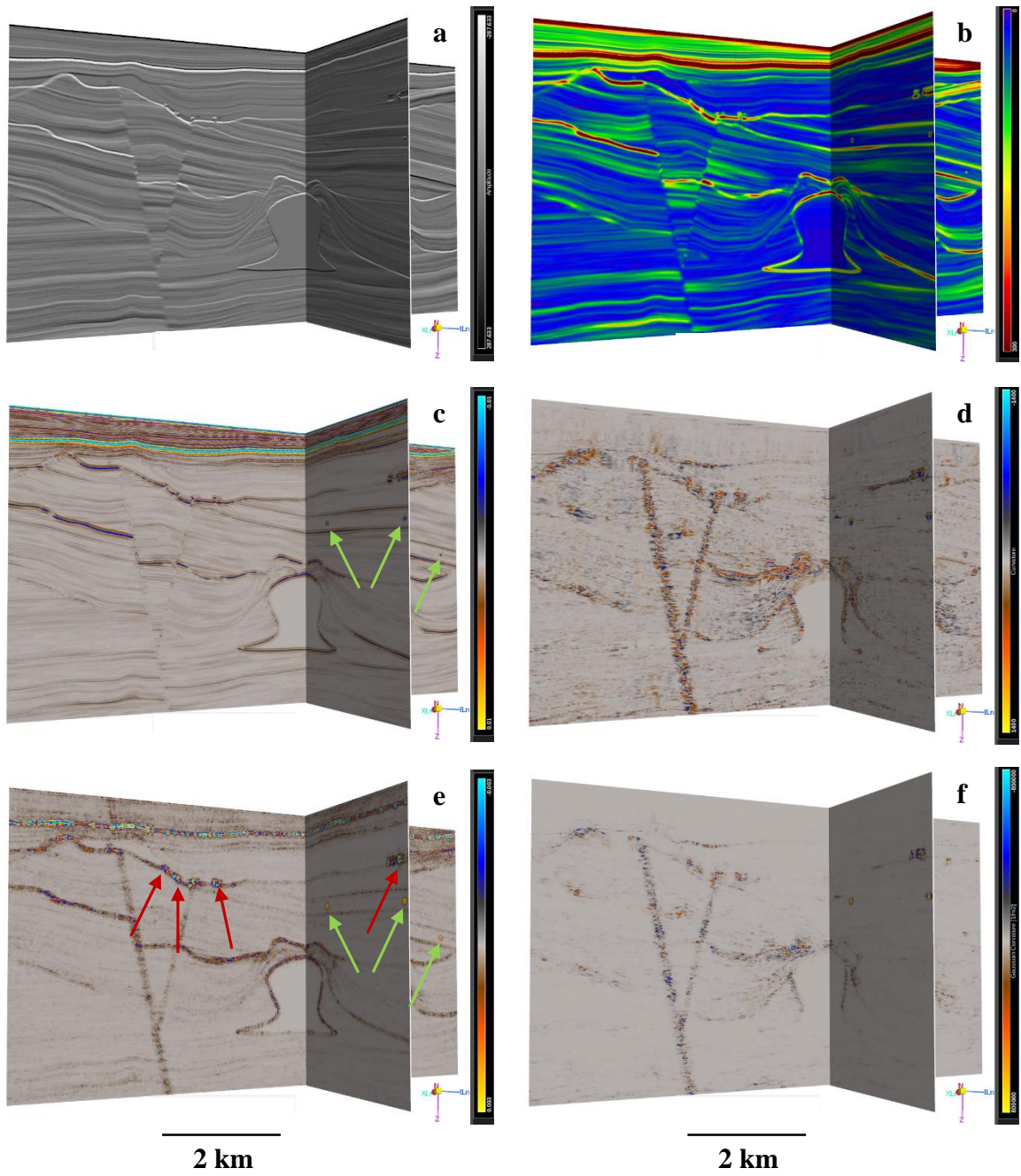

Figure 6. Volumetric curvatures of seismic envelope, 3D views, synthetic model

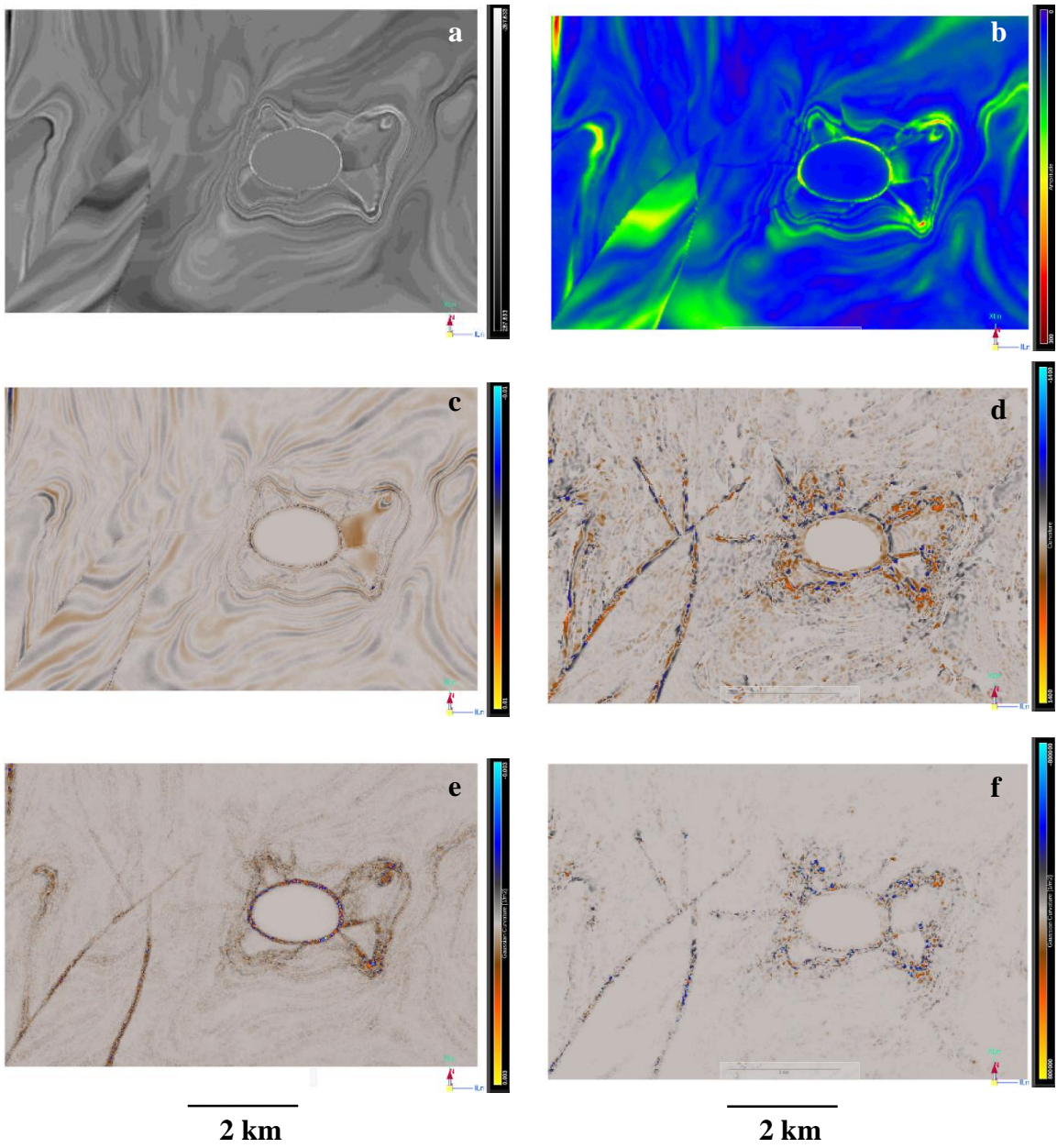

**2 km**

Figure 7. Volumetric curvatures of seismic envelope, horizontal slice views, synthetic model

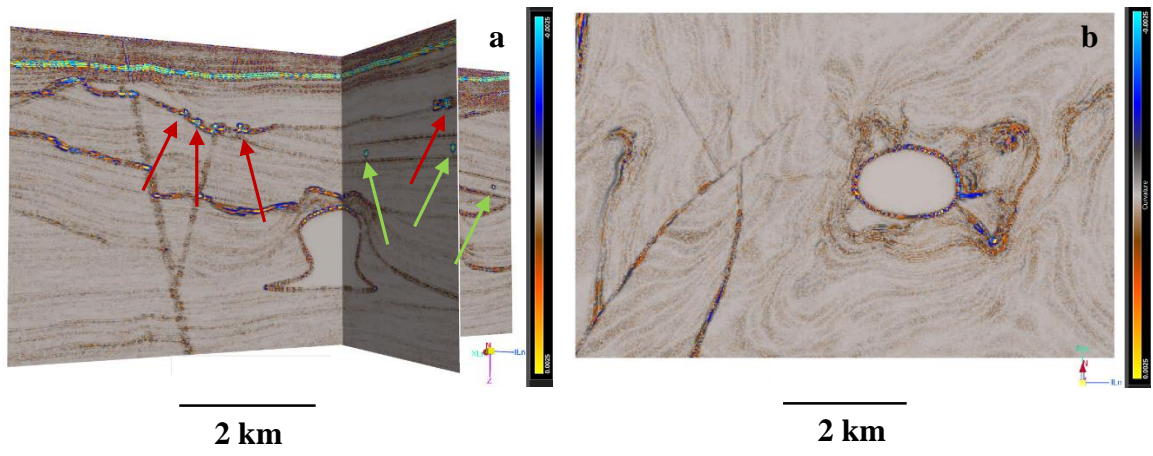

Figure 8. Scalar hypersurface curvature of seismic envelope, synthetic model

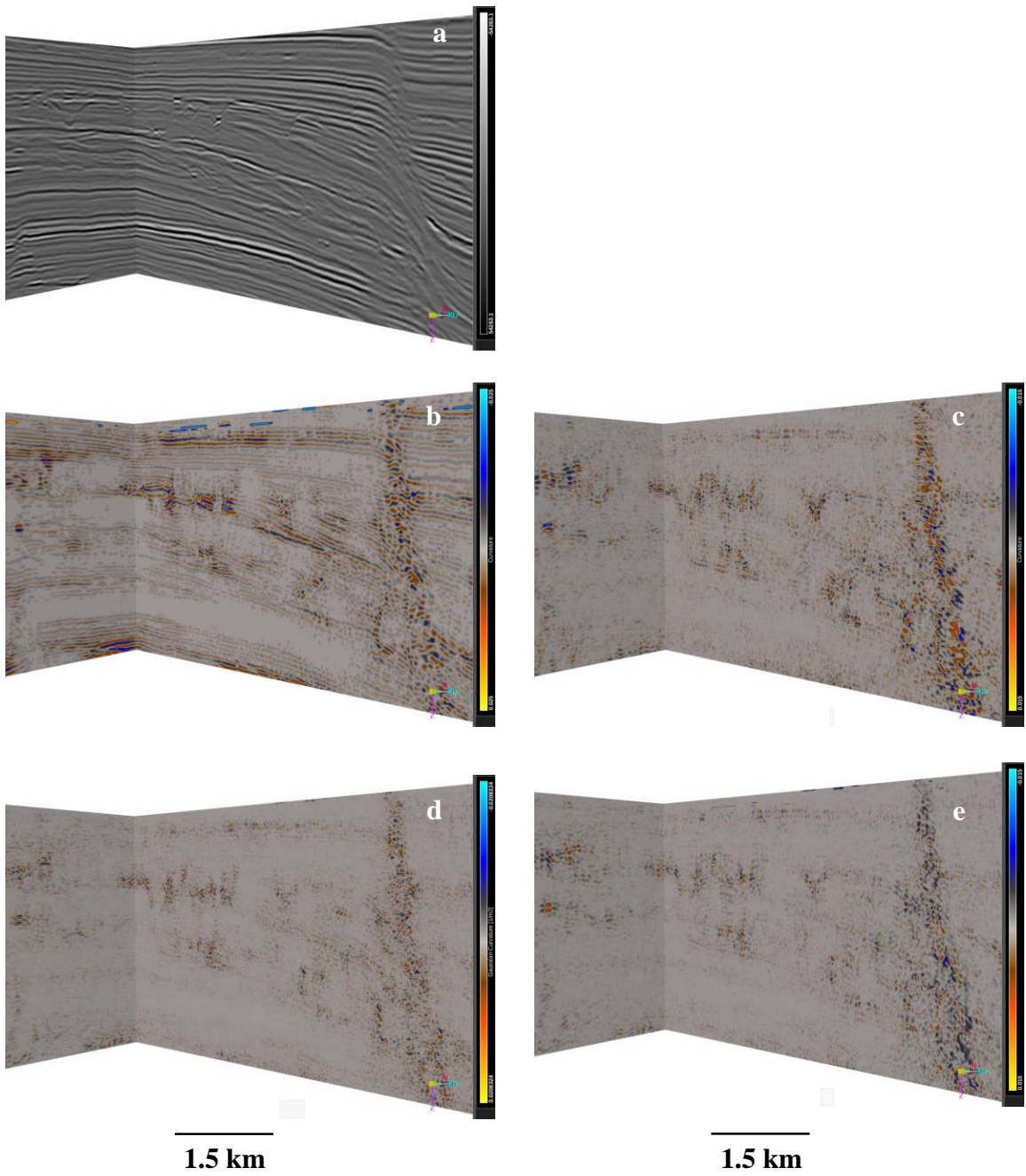

**1.5 km**          **1.5 km**

Figure 9. Volumetric curvatures of seismic amplitude, 3D views, Parihaka dataset

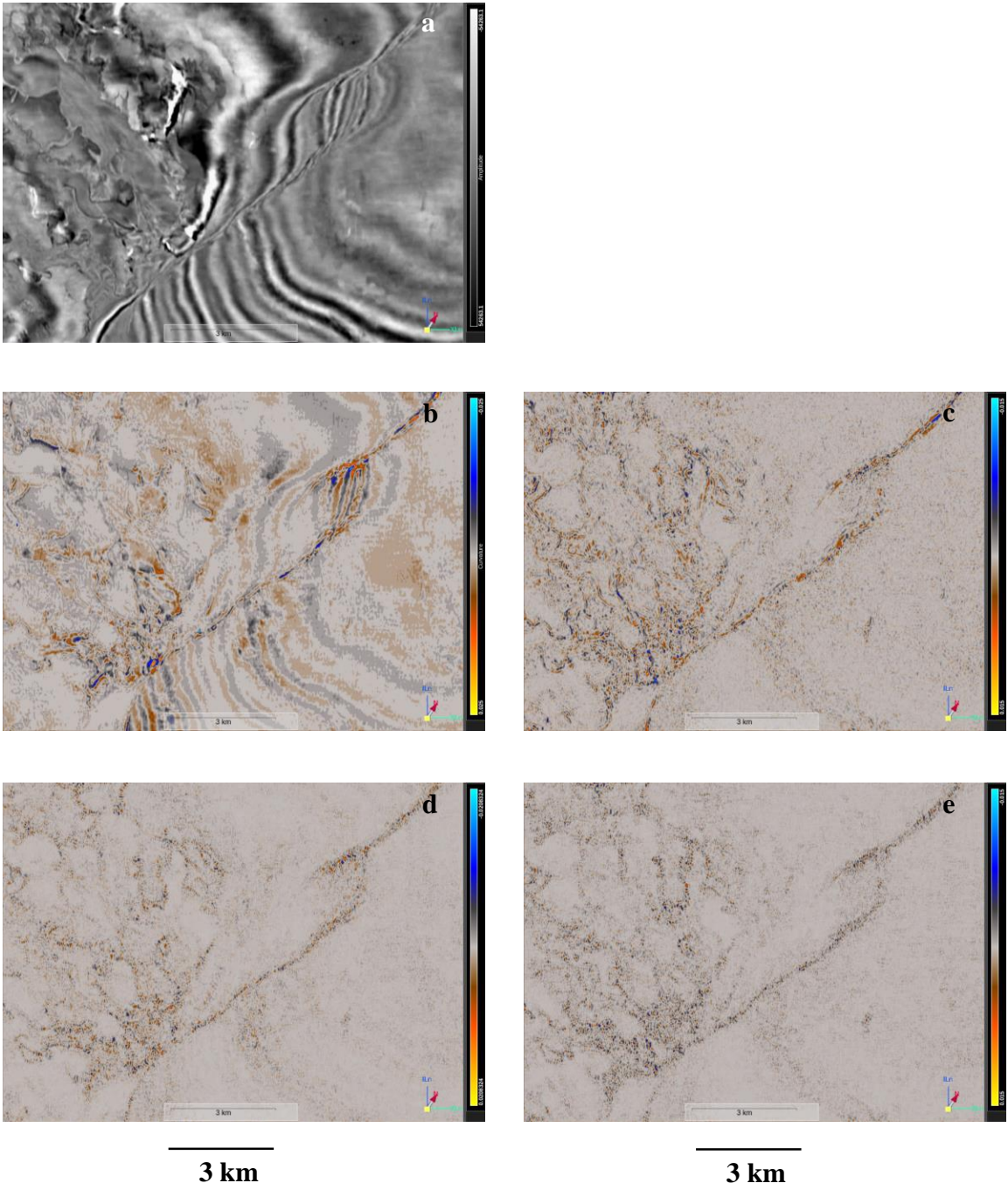

Figure 10. Volumetric curvatures of seismic amplitude, horizontal slice views, Parihaka dataset

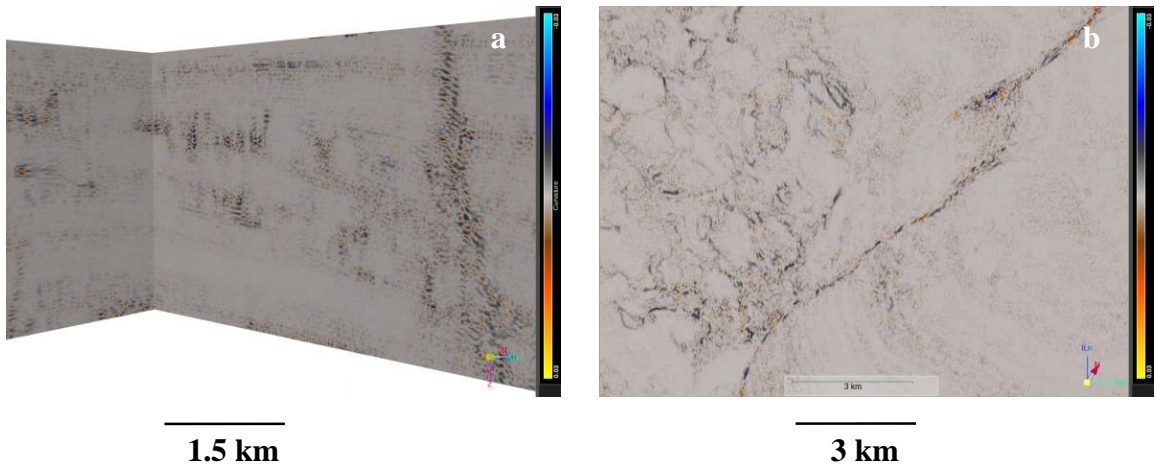

**1.5 km**          **3 km**

Figure 11. Scalar hypersurface curvature of seismic amplitude, Parihaka dataset

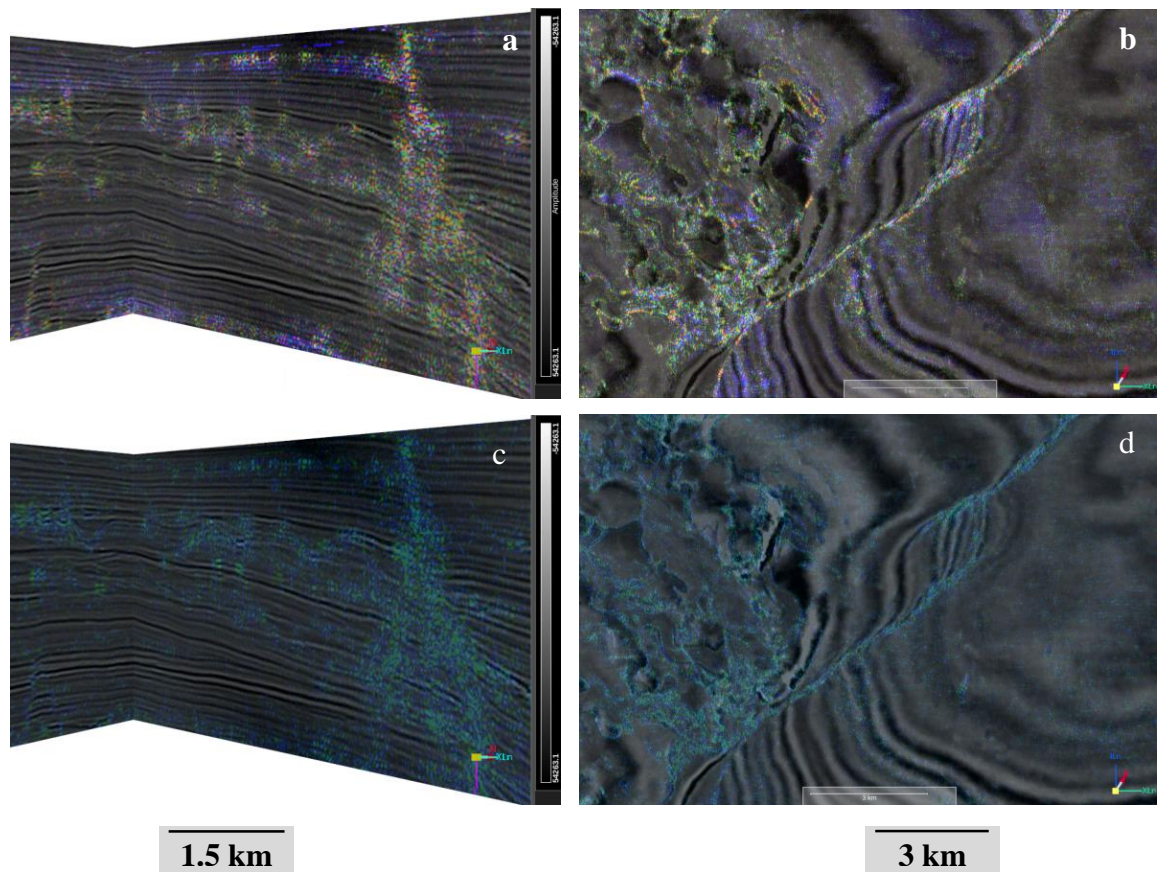

**1.5 km**   **3 km**

Figure 12. Seismic image field, co-rendered with three effective hypersurface curvatures (a, b) and two effective dip-based curvatures (c, d) of Parihaka dataset: Blue, red and green colors represent the absolute values of the mean, scalar, and Gaussian curvatures, respectively

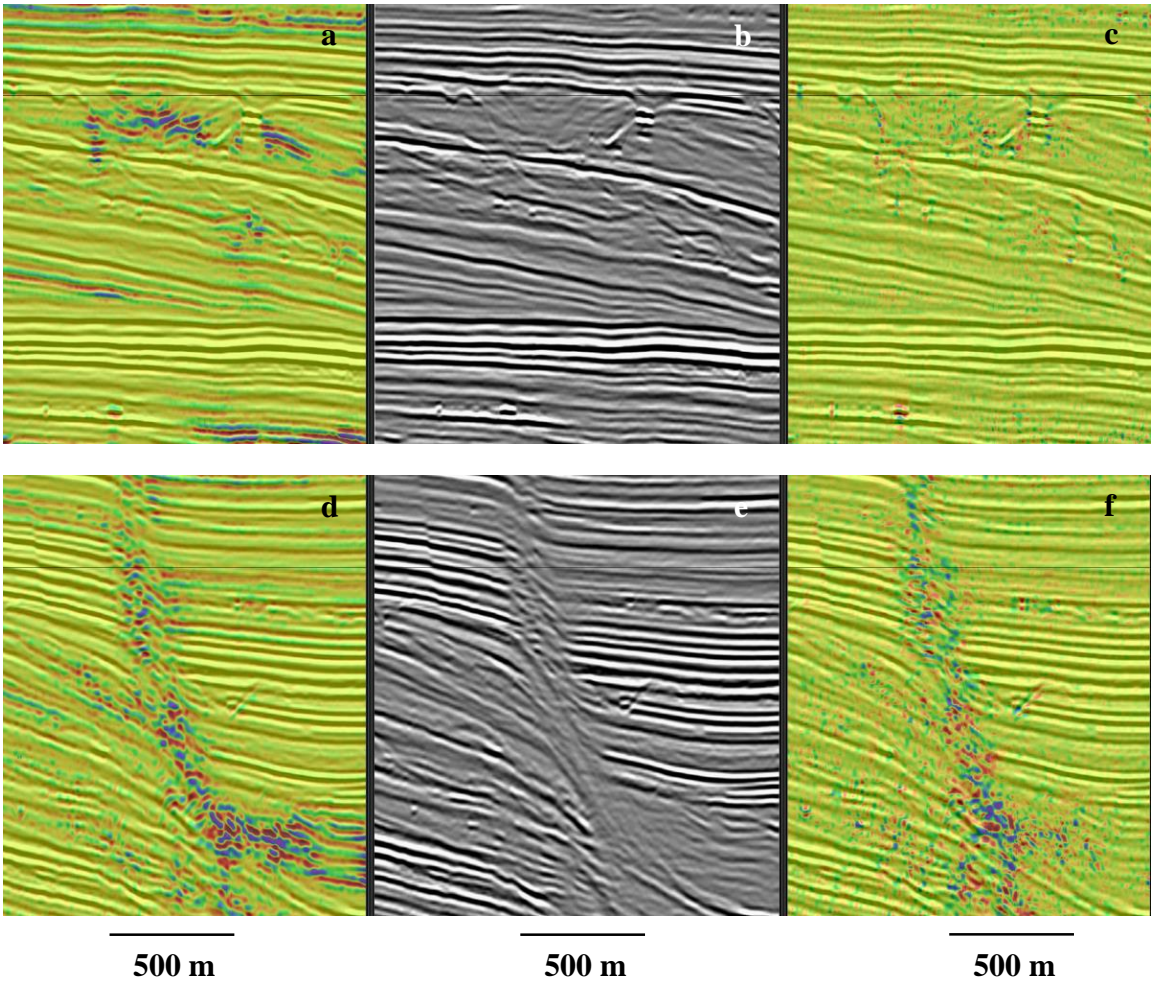

500 m       500 m       500 m

Figure 13. Channel and fault of Parihaka dataset: Seismic signal and mean curvature

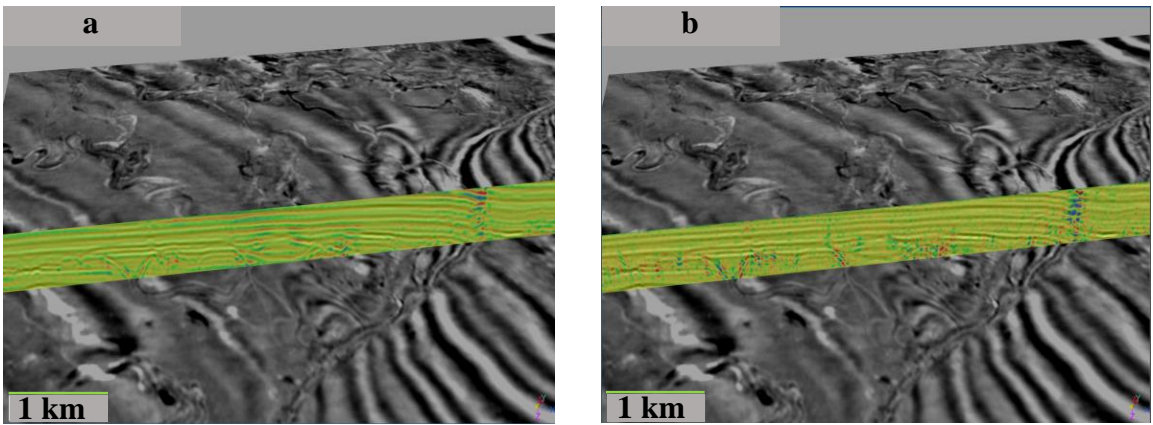

Figure 14. Mean curvature: horizontal slice and inline section, Parihaka dataset

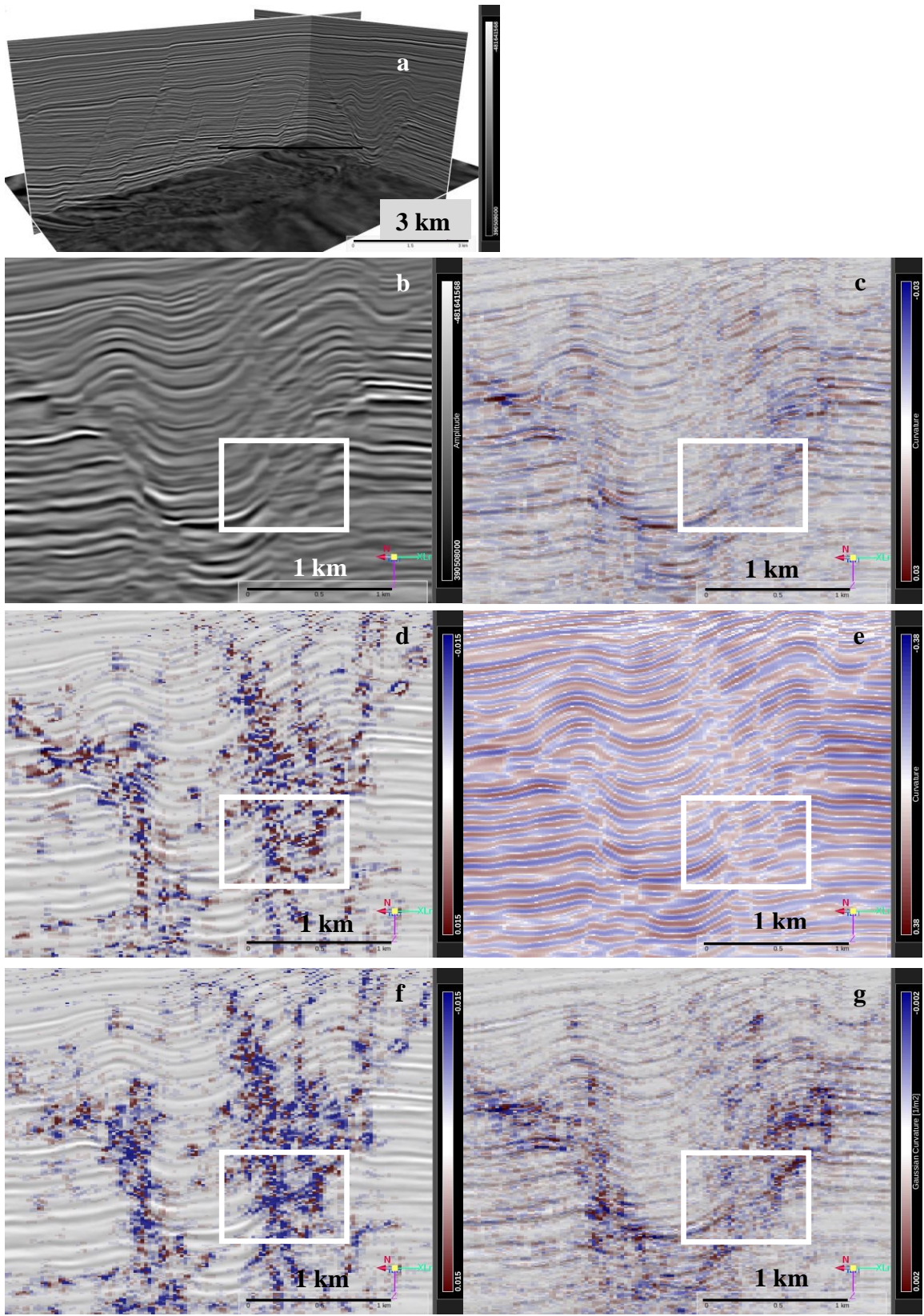

Figure 15. Strike-slip fault zone on inline section, Clyde dataset

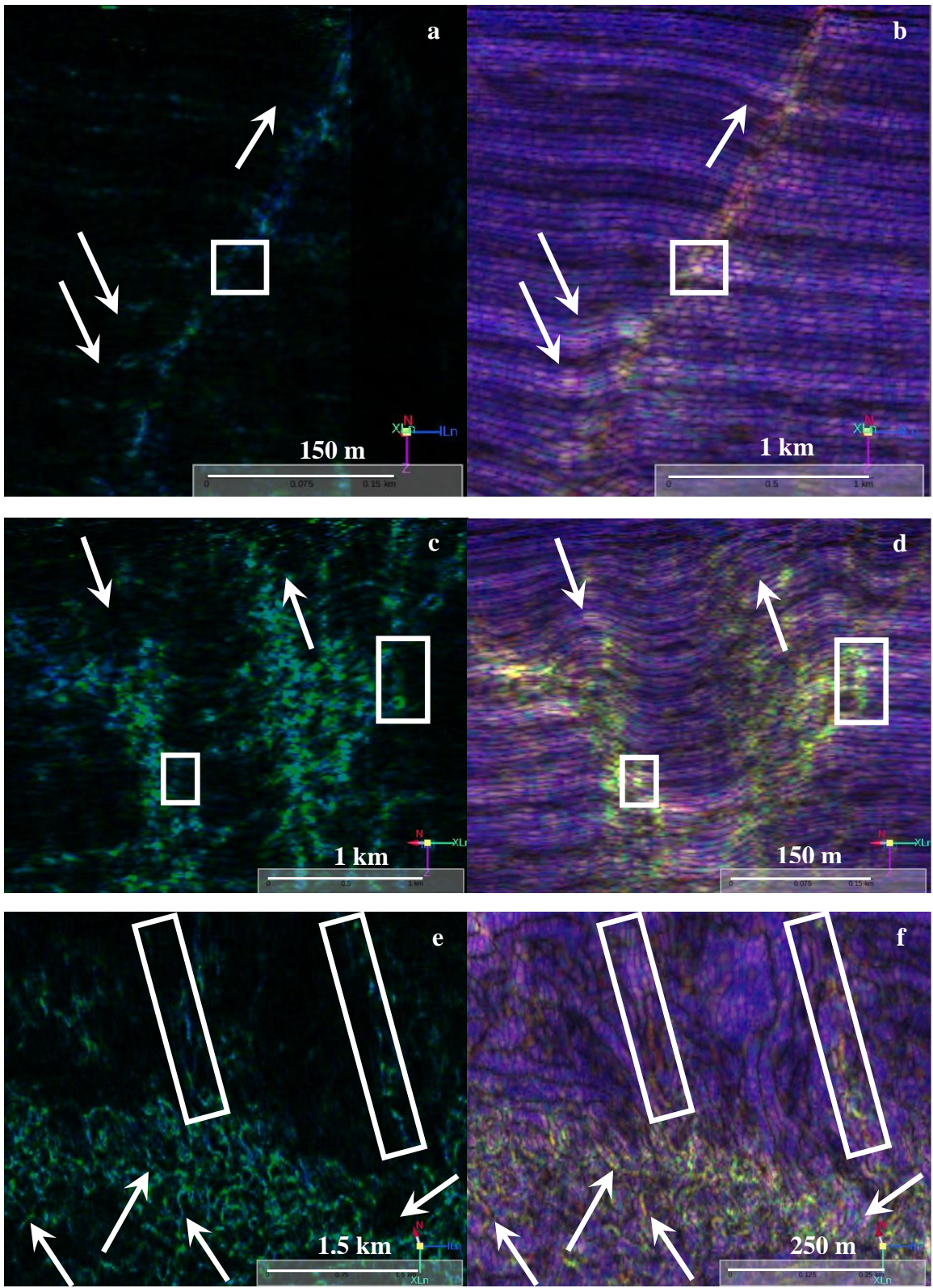

Figure 16. Combining curvature attributes in RGB display, Clyde dataset

**TABLES**

Table 1. Synthetic model: Approximate depth ranges and

velocity distributions per stratigraphic sequence

| Stratigraphic sequence | Approximate depth range, km | Velocity distribution | |
|---|---|---|---|
| | | Type | Range, km/s |
| 1 | $0 - 0.56$ | Triangular | $2.16 - 2.36$ |
| 2 | $0 - 1.40$ | Uniform | $3.4 - 3.5$ |
| 3 | $0.5 - 2.4$ | Uniform | $3.5 - 3.6$ |
| 4 | $0.75 - 2.6$ | Uniform | $3.5 - 3.6$ |
| 5 | $1.5 - 2.6$ | Uniform | $3.6 - 3.7$ |
| 6 | $1.5 - 3.7$ | Uniform | $3.6 - 3.7$ |
| 7 | $2.6 - 4.8$ | Uniform | $3.7 - 3.8$ |
| 8 | $4.8 - 5.0$ | Uniform | $3.7 - 3.8$ |

Table 2. Surface types vs. signs of most positive and most negative curvatures

(surface types below the diagonal do not exist)

| $k_{pos}$ & $k_{neg}$ | $k_{pos} < 0$ | $k_{pos} = 0$ | $k_{pos} > 0$ |
|---|---|---|---|
| $k_{neg} < 0$ | Bowl | Synform | Saddle |
| $k_{neg} = 0$ | | Plane | Antiform |
| $k_{neg} > 0$ | | | Dome |

**LIST OF TABLES**

**LIST OF FIGURES**

Figure 1. Modeling structure in the geological space and properties in UVT-transformed space.

    a.   Structural model comprising 14 horizons (only seven are shown), 22 faults, and a salt body.

    b.   Relative geological time computed from the structural data (faults and horizons) over the entire geological space.

    c.   Geostatistical simulation of density in the depositional space. Density increases with depth, is strongly correlated laterally, and shows more rapid variations in the vertical direction. A channel with contrasting density is visible in the upper part of the volume.

    d.   Geostatistical simulation of density in one of the horizontal layers containing the channel. Both channel and point bar facies are simulated with densities that contrast with the general, smoother background density of the layer.

Figure 2.  Combining structure and geological features in geological space

    a.   Geostatistical simulation of density, transported to the geological space via the UVT Transform. Density is now strongly correlated laterally inside each stratigraphic sequence, and discontinuous across faults and unconformities, following the structure. The contrasting channel facies are visible in the upper part of the model. Punctual density anomalies added to one of the layers below are not visible.

b. Geostatistical simulation of velocity in the geological space. Velocity is strongly correlated laterally in each stratigraphic layer and discontinuous across faults and unconformities, following the structure. Contrasting velocities can be seen in the channel.

c. Impedance volume in the geological space, obtained from density and velocity volumes. Impedance values are blurred at fault and unconformity locations to avoid creating excessively sharp contrasts.

d. Seismic reflectivity generated in the geological space from convolving impedance traces with a wavelet. Horizons are overlaid as colored lines. The density anomalies introduced in one of the upper layers are visible as point diffractors.

Figure 3. Volumetric curvatures of velocity field, 3D views, synthetic model:

a. Original velocity field

b. Smoothed velocity filed

c. Mean curvature of hypersurface

d. Mean curvature of 2D surface (dip-based mean curvature)

e. Gaussian curvature of hypersurface

f. Gaussian curvature of 2D surface (dip-based Gaussian curvature)

In the plot indicating the mean hypersurface curvature, fine variations and stratigraphic discontinuities inside the layers are highlighted with a higher resolution than those in the plot indicating the dip-based mean curvature. With the hypersurface curvatures, the boundary of the salt body is completely recovered, even in parts where its dip is close to the dip of the background stratigraphic layers. The Gaussian curvature of the hypersurface highlights discontinuities in the seismic image.

Figure 4. Volumetric curvatures of velocity field, horizontal slice views, synthetic model:

a. Original velocity field

b. Smoothed velocity filed

c. Mean curvature of hypersurface

d. Mean curvature of 2D surface (dip-based mean curvature)

e. Gaussian curvature of hypersurface

f. Gaussian curvature of 2D surface (dip-based Gaussian curvature)

The dip-based mean curvature is noisier and less continuous than the hypersurface mean curvature. It fails to highlight the whole boundary of the salt body, and does not enhance the channel enough to individualize it as a specific geological object. The Gaussian curvature of the hypersurface highlights discontinuities in the seismic image in a more contrasted way than that of the dip-based surface.

Figure 5. Scalar hypersurface curvature of velocity field, synthetic model:

a. 3D view

b. Horizontal slice view

The scalar hypersurface curvature combines characteristics from the mean and Gaussian curvatures.

Figure 6. Volumetric curvatures of seismic envelope, 3D views, synthetic model:

a. Original seismic image field

b. Smoothed seismic envelope field

c. Mean curvature of hypersurface

d. Mean curvature of 2D surface (dip-based mean curvature)

e. Gaussian curvature of hypersurface

f. Gaussian curvature of 2D surface (dip-based Gaussian curvature)

The channel feature above the stratigraphic unconformity in the top third of the model stands out as a set of bright spots where the cross-section intersects the object (red arrows). The point diffractors, added as small, dense spheres in the density volume, appear in the mean and Gaussian curvatures of the seismic envelope hypersurface as bright spots on a clean background (green arrows), whereas these objects appear on the level of noise in the dip-based curvatures.

Figure 7. Volumetric curvatures of seismic envelope, horizontal slice views, synthetic model:

a. Original seismic image field

b. Smoothed seismic envelope field

c. Mean curvature of hypersurface

d. Mean curvature of 2D surface (dip-based mean curvature)

e. Gaussian curvature of hypersurface

f. Gaussian curvature of 2D surface (dip-based Gaussian curvature)

The hypersurface mean curvature is much more continuous along geological structures compared to the dip-based mean curvature. Discontinuities of the hypersurface Gaussian curvature, such as faults and stratigraphic unconformities, are sharper and less noisy than those of the dip-based Gaussian curvature.

Figure 8. Scalar hypersurface curvature of seismic envelope, synthetic model

a.   3D view

b.   Horizontal slice view

The scalar hypersurface curvature retains a strong continuity along stratigraphic structures and highlights discontinuities.

Figure 9. Volumetric curvatures of seismic amplitude, 3D views, Parihaka dataset:

a.   Original seismic image field

b.   Mean curvature of hypersurface

c.   Mean curvature of 2D surface (dip-based mean curvature)

d.   Gaussian curvature of hypersurface

e.   Gaussian curvature of 2D surface (dip-based Gaussian curvature)

The hypersurface mean curvature shows better continuity along reflectors than the dip-based curvature, highlighting horizons and transition zones. The dip-based curvature effectively ignores the second vertical derivative of a scalar field, which is not optimal for horizon detection. The Gaussian curvature of the hypersurface is less noisy than the dip-based Gaussian curvature.

Figure 10. Volumetric curvatures of seismic amplitude, horizontal slice views, Parihaka dataset:

a. Original seismic image field

b. Mean curvature of hypersurface

c. Mean curvature of 2D surface (dip-based mean curvature)

d. Gaussian curvature of hypersurface

e. Gaussian curvature of 2D surface (dip-based Gaussian curvature)

The hypersurface mean curvature demonstrates better continuity; it highlights channel edges and a relay area between two faults. The Gaussian curvature of the hypersurface shows better definition of the faults.

Figure 11. Scalar hypersurface curvature of seismic amplitude, Parihaka dataset

a. 3D view

b. Horizontal slice view

The hypersurface scalar curvature clearly highlights faults and edges of channels with a good definition.

Figure 12. Seismic image field, co-rendered with three effective hypersurface curvatures (a, b) and two effective dip-based curvatures (c, d) of Parihaka dataset: Blue, red and green colors represent the absolute values of the mean, scalar, and Gaussian curvatures, respectively

a. Effective hypersurface curvatures, 3D view

b. Effective hypersurface curvatures, horizontal slice view

c. Effective dip-based curvatures, 3D view

d. Effective dip-based curvatures, horizontal slice view

For the hypersurface curvatures (upper plots), the mean curvature, mapped to the blue channel, highlights the geological structures which are mainly discontinuous in the vertical direction. Its Gaussian curvature, mapped to the green channel, mostly picks discontinuities such as faults and edges of channels. The scalar curvature, mapped to the red channel, shows a mix of structure and discontinuous characteristics. The dip-based surface curvatures have no scalar curvature, so the red color is absent in the lower plots. The hypersurface effective curvatures have a better definition than the corresponding dip-based curvatures, and therefore, the RGB plot for the effective hypersurface curvatures is much sharper than that for the dip-based surfaces. Furthermore, mapping the additional scalar hypersurface curvature to the red channel enhances areas where all three effective curvatures are high by making them white.

Figure 13. Channel and fault of Parihaka dataset: Seismic signal and mean curvature

a.  Channel, mean curvature of hypersurface

b.  Channel, seismic amplitude

c.  Channel, mean curvature of 2D surface (dip-based mean curvature)

d.  Fault, mean curvature of hypersurface

e.  Fault, seismic amplitude

f.  Fault, mean curvature of 2D surface (dip-based mean curvature)

The internal structure of the large channel, which cannot be seen in the (input) seismic image, is clearly visible with the hypersurface mean curvature. This is not so for the dip-based curvature. In the proximity of the fault, the hypersurface mean curvature shows better continuity even in the low energy regions: The internal structure of the fault area, which is blurred in the seismic image, is clearly visible with the hypersurface mean curvature, which is not the case with the dip-based curvature.

Figure 14. Mean curvature: horizontal slice and inline section, Parihaka dataset

a.  Mean curvature of hypersurface

b.  Mean curvature of 2D surface (dip-based mean curvature)

Channel features along the horizontal slice are correlated with the internal structure along the vertical slice, which is highlighted much more strongly and clearly by the hypersurface mean curvature than by the dip-based mean curvature.

Figure 15. Strike-slip fault zone on inline section, Clyde dataset

a.  Original seismic image, 3D view

b.  Seismic amplitude, inline section

c.   Amplitude with hypersurface scalar curvature overlay

d.  Amplitude with dip-based mean curvature overlay

e.  Hypersurface mean curvature

f.  Amplitude with dip-based Gaussian curvature overlay

g.  Amplitude with hypersurface Gaussian curvature overlay

Figure 16. Combining curvature attributes in RGB display, Clyde dataset

a. Dip-based Gaussian (green) and mean (blue) curvatures, crossline section

b. Hypersurface scalar (red), Gaussian (green), and mean (blue) curvatures

c. Dip-based Gaussian (green) and mean(blue) curvatures, inline section

d. Hypersurface scalar (red), Gaussian (green), and mean (blue) curvatures

e. Dip-based Gaussian (green) and mean (blue) curvatures, depth slice

f. Hypersurface scalar (red), Gaussian (green), and mean (blue) curvatures